\documentclass[a4paper,fleqn,usenatbib]{mnras}
\usepackage{newtxtext,newtxmath}
\usepackage[T1]{fontenc}
\usepackage{ae,aecompl}
\usepackage{graphicx}    % Including figure files
\usepackage{amsmath}    % Advanced maths commands
\usepackage{amssymb}    % Extra maths symbols
\usepackage{pdflscape}
\usepackage{comment}
\usepackage{subcaption}
\usepackage{hyperref}
\usepackage{breakurl}
\hypersetup{pdftitle= Rev-1 NGC 6300}

\title[Nuclear and Circumnuclear Properties NGC 6300]{Probing Nuclear and Circumnuclear Properties of NGC 6300 using X-ray Observations}

\author[Jana et al.]{
Arghajit Jana,$^{1}$\thanks{\href{mailto:argha@prl.res.in}{argha@prl.res.in}}
Arka Chatterjee,$^{2}$\thanks{\href{mailto:arkachatterjee@bose.res.in}{arkachatterjee@bose.res.in}}
Neeraj Kumari,$^{1,3}$
Prantik Nandi,$^{2}$
Sachindra Naik$^{1}$
\newauthor
and Dusmanta Patra$^{4}$
\\
$^{1}$Astronomy \& Astrophysics Division, Physical Research Laboratory, 
      Navrangpura, Ahmadabad, 380009, India\\
$^{2}$Department of Astrophysics \& Cosmology, S. N. Bose National Centre for 
      Basic Science, Block-JD, Sector-III, Salt Lake, Kolkata, 700106, India\\
$^{3}$Department of Physics, Indian Institute of Technology, 
      Gandhinagar, 382355, Gujarat, India\\
$^{4}$Indian Centre for Space Physics, Garia Staion Road, 
      Kolkata, 7000084, India\\}

\date{Accepted XXX. Received YYY; in original form ZZZ}

\pubyear{2020}

\begin{document}

\label{firstpage}
\pagerange{\pageref{firstpage}--\pageref{lastpage}}
\maketitle

\begin{abstract}

We present the results obtained from a detailed X-ray timing and spectral analysis of Seyfert~2 galaxy
NGC~6300 by using observations with the {\it Suzaku}, {\it Chandra} and {\it NuSTAR} observatories 
between 2007 and 2016. We calculate variance, rms fractional variability of the source in different 
energy bands and find variabilities in various energy bands. Spectral properties of the source are 
studied by using various phenomenological and physical models. The properties of the Compton clouds, 
reflection, Fe K$\alpha$ line emission and soft X-ray excess are studied in detail. Several physical 
parameters of the source are extracted and investigated to establish the presence/absence of any 
correlation between them. We also investigate the nature of the circumnuclear `torus' and find that 
the torus is not uniform, rather clumpy. The observed changes in the line-of-sight column density 
can be explained in terms of transiting clouds. The iron line emitting region is found to be different 
in the different epoch of observations. We also observe that the torus and the nucleus independently 
evolve over the years.

\end{abstract}


%%%%%%%%%%%%%%%%%%%%%%%%%%%%%%%%%%%%%%%%%%%%%%%%%%

%%%%%%%%%%%%%%%%% BODY OF PAPER %%%%%%%%%%%%%%%%%%

\section{Introduction}

Active galactic nuclei (AGNs) are the most energetic persistent objects in the universe. The AGNs are 
powered by the accreting supermassive black holes (SMBH) which reside at the centre of each galaxy 
\citep{Rees1984}. The matter from the surrounding medium is accreted in the form of a geometrically 
thin, optically thick accretion disc around the blackhole \citep{SS73}, also known as the standard disc. 
The accretion disc, in case of AGNs, predominately radiates in the UV/optical wavebands and creates 
the so-called `big-blue-bump' in the broadband spectral energy distribution (SED). The X-rays, on the 
other hand, are emitted from a Compton cloud located within a few tens of Schwarzschild radii around 
the central engine \citep{HM91, Fabian2015}. The X-rays can be produced by the inverse-Compton scattering
of the UV/optical photons originating from the accretion disc. The hard X-ray photons are reflected at
the relatively cold matter and produce Fe fluorescent line \citep{GF1991, Matt1991}. Thus, along with 
the primary X-ray continuum, a reflection hump at $\sim$15-30~keV energy range, and a Fe fluorescent 
line at $\sim$6.4~keV are observed in the X-ray spectrum of AGNs. An additional soft X-ray (< 2~keV) 
component, known as `soft excess' is often observed \citep{Halpern84, Arnaud1985, Singh1985} in the 
AGNs spectra. The `soft excess' could have a completely different origin than the primary continuum, 
and is often associated with the host galaxy. However, more recent studies \citep{C2006, SD2007} 
indicate that the `soft-excess' could be generated via reflection mechanism or due to the variable 
nature of the hydrogen column density ($N_H$) along the line of sight. Inverse Comptonization
by warm and optically thick region \citep{Mehdipour2011} and relativistic blurred reflection of the hot 
coronal photons by the inner disc \citep{Nardini11} are also considered to be the origin of the soft excess .
\citet{Lohfink2012} suggested that hot corona to be the origin of the soft excess. \citet{Garcia2019}
explored both the possibilities, i.e. relativistic reflection as well as Comptonization form warm corona 
in Mrk 509. The AGNs also produce powerful relativistic jets which are best observed in radio bands. 
About $\sim$15 per cent AGNs show relativistic jets, leading to a classification of radio-loud and radio-quiet
AGNs \citep{Urry1995}. Recently, \citet{Panessa2016} reported that from the total AGN population,
about $\sim 7-10$ per cent galaxies are the radio galaxies.

AGNs are classified as type-I or type-II based on the observation of broad line emission (originate
in the broad line emitting region or BLR) or narrow line emission (originate from narrow line emitting region
or NLR) in the optical wavebands \citep{Antonucci1993}. In type-I AGN, both broad lines and narrow lines are
observed, while in type-II AGN, only narrow lines are observed in the optical spectra. Observation of broad 
lines in the polarized light in NGC~1068 revealed that the BLR is obstructed by an absorbing dusty `torus' 
surrounding the AGN \citep{Antonucci1985}. This leads to the unified model for AGN, where the classification 
is due to the orientation effect. The unified model of AGN from the optical wavebands transforms into the X-ray 
wavebands as the hydrogen column density ($N_H$) in the line-of-sight of the absorbing material. The type-I AGNs 
are observed in unobstructed way with $N_H < 10^{22}$ cm$^{-2}$, while the type-II AGNs are observed through the 
obstruction where $N_H > 10^{23}$ cm$^{-2}$. In the case of type-II AGNs, a dusty torus around the accretion 
disc is considered as the major absorbing medium along the line-of-sight of the observer although 
the possibility of contribution from NLRs and BLRs towards total $N_H$ can not be ignored. The obscuring torus 
is characterized by its hydrogen column density ($N_H$). If $N_H > 1.5 \times 10^{24}$ atoms $cm^{-2}$, the torus 
is considered as Compton thick; otherwise, the torus is Compton thin. Generally, photons with energy $< 2$ keV 
suffers absorption and above $\sim 2$~keV, the unabsorbed spectrum is expected. The absorption of X-rays 
increases as the column density of absorbing material increases. In the case of torus geometry, flux suppression 
in the energy range below 10~keV gets flatten off in case of Compton-thick source, i.e. $N_H > 1.5 \times 10^{24}$ 
cm$^{-2}$, as there is always reflected flux contribution from the torus irrespective of the obscuration level of 
the torus \citep{Brightman2011}.

Seyfert~2 galaxies are the radio-quiet type-II AGNs \citep{Netzer2013}. In this case, a dusty torus 
surrounds the circumnuclear region. It is believed that the `torus' is located far away (a few parsecs) from 
the nucleus. In general, the nature of the torus does not change over the years. Although, for several AGNs, 
it is observed that the opacity of the torus changes from Compton-thin to Compton-thick and vice-versa in a 
timescale ranging from months to years \citep{Risaliti2002, Matt2003}. This type of AGNs are known as 
changing-look AGN. The variation of column density is believed to be occurred due to the transition of cloud 
along the line-of-sight. The location of these transiting clouds commensurate with the outer region of the BLR 
or dusty inner torus and sometimes with the inner BLR \citep{Markowitz2014}. \citet{HG2015} showed that 
Compton-thin and changing-look AGNs show more variabilities than the Compton-thick AGN.

NGC~6300 is a nearby Seyfert~2 galaxy with $z=0.0037$ \citep{Meyer2004}. It is located at R.A. = $17^h$ $16^m$ 
$59.473^s$; DEC = $=-62$\textdegree $49'$ $13''.98$ \citep{Skrutskie2006}. It is a ring, barred spiral galaxy
and classified as SBb-type galaxy from its morphology. NGC~6300 was observed a few times in X-ray bands with the 
RXTE \citep{Leighly99}, BeppoSAX \citep{Matsumoto2004} and XMM-Newton \citep{Guainazzi2002}, in February 1997,
August 1999 and March 2001, respectively. From the variability studies with the XMM-Newton, the mass of NGC~6300
was estimated to be $\sim 2.8 \times 10^5$ $M_{\odot}$. The $M_{BH}-\sigma$ relation yields the mass of NGC~6300 
to be $10^7$ $M_{\odot}$. However, with various uncertainties, the mass is estimated to be $< 10^7$ $M_{\odot}$
\citep{Awaki2005}. \citet{Khorunzhev2012} estimated the mass of the BH to be $10^{7.59}$ $M_{\odot}$ from mass-bulge
luminosity correlation. The NIR study of the molecular radial velocity yields the mass of the BH as $< 6.25 \times 
10^7$ $M_{\odot}$ \citep{Gaspar2019}.

In this article, we study the Seyfert~2 galaxy NGC~6300 by using observations between 2007 \& 2016 at five epochs 
(2007, 2009, 2013, January 2016 \& August 2016) with the {\it Suzaku}, {\it Chandra}, and {\it NuSTAR} observatories. 
We investigate the characteristics of the nucleus as well as the nature of the circumnuclear torus. In \S 
\ref{sec:observation}, we briefly discuss the observations and data reduction processes. In \S \ref{sec:analysis}, 
we present timing and spectral analysis methods and the corresponding results. In \S \ref{sec:discussion}, we discuss 
our findings. Finally, we draw our conclusions in \S \ref{sec:conclusion}.

\section{Observation and Data Reduction}
\label{sec:observation}

\begin{table}
\caption{Log of observations of NGC~6300.}
\label{tab:1}
\begin{tabular}{lcccc}
\hline
ID&Date & Obs. ID & Instrument & Exposures\\
&(yyyy-mm-dd) & &  &(ks)\\
\hline
S1 & 2007-10-17 & 702049010 & {\it Suzaku} & 82.5 \\
C1 & 2009-06-03 & 10289 & {\it Chandra}/ACIS & 10.2 \\
C2 & 2009-06-07 & 10290 & {\it Chandra}/ACIS & 9.8  \\
C3 & 2009-06-09 & 10291 & {\it Chandra}/ACIS & 10.2 \\
C4 & 2009-06-10 & 10292 & {\it Chandra}/ACIS & 10.2 \\
C5 & 2009-06-14 & 10293 & {\it Chandra}/ACIS & 10.2 \\
N1 & 2013-02-25 & 60061277002 & {\it NuSTAR} & 17.7 \\
N2 & 2016-01-24 & 60261001002 & {\it NuSTAR} & 20.4 \\
N3 & 2016-08-24 & 60261001004 & {\it NuSTAR} & 23.5 \\
\hline
\end{tabular}
\end{table}

We searched and acquired the publicly available archival data of NGC~6300 from {\it Suzaku}, {\it Chandra} and {\it Nustar} 
observatories by using HEASARC\footnote{\url{http://heasarc.gsfc.nasa.gov/}}.

\label{sec:suzaku}
\subsection{Suzaku}

NGC~6300 was observed with {\it Suzaku} on 2007 October 17 (Obs ID: 702049010). The {\it Suzaku} observatory consisted
of two sets of instruments: the X-ray Imaging Spectrometer (XIS) \citep{Koyama2007} and the Hard X-ray Detector (HXD) 
\citep{Takahashi2007}. There are four units of XIS among which XIS-0, XIS-2, and XIS-3 were front-side-illuminated CCDs
(FI-XISs), while XIS-1 was back-side-illuminated one (BI-XIS). The HXD was a non-imaging Instrument consisting of Si PIN
photo-diodes and GSO scintillation counters. We followed\footnote{\url{http://heasarc.gsfc.nasa.gov/docs/suzaku/analysis/abc/}} 
standard procedures and recommended screening criteria while extracting {\it Suzaku}/XIS spectra and light-curves. We 
reprocessed the event files by using the latest calibration data files \footnote{\url{http://www.astro.isas.jaxa.jp/suzaku/caldb/}},
released on 2014-02-03, through the software package {\tt FTOOLS 6.25}. We chose a circular region with a radius 
of 210 arcsecs and source coordinates as the centre, for source extraction. The background spectra were extracted from 
source-free regions by selecting a circular region of 210 arcsecs radius. With the {\tt xisrmfgen} and {\tt xisarfgen} 
task, we generated Response matrices and ancillary response files, respectively. As XIS-2 was not operational, the data 
from the other three XISs are used in the present analysis. The $2-10$~keV front-illuminated XIS-0 and XIS-3 spectra are 
co-added using the {\tt addascaspec} task, whereas the $0.5-10$ keV back-illuminated XIS-1 spectrum was treated separately. 
The XIS spectra in $1.6-2$~keV range was ignored due to the presence of known Si edge. All the spectra are re-binned to 
achieve $>$20 counts per channel bins by using the {\tt grppha} task. For {\it Suzaku} HXD/PIN spectra, cleaned event files 
were generated using the {\tt aepipeline} task. With the {\tt hxdpinxbpi} task, the deadtime corrected source and background
spectra were generated. The background spectra included non-X-ray background (nxb; \citet{Fukazawa2009}) and the simulated 
cosmic X-ray background (cxb; \citet{Gruber1999}). We used HXD/PIN spectrum in $15-40$~keV energy range in our analysis.

\label{sec:chandra}
\subsection{Chandra}

The {\it Chandra}/ACIS observed NGC~6300 five times between 2009 June 03 and 2009 June 14. We summarized these observations 
in Table \ref{tab:1}. All the observations were carried out in Faint data mode. The data were processed with the Chandra 
Interactive Analysis of Observations tools (CIAO v.4.11\footnote{\url{https://cxc.harvard.edu/ciao/download/}};\citep{FMA2006,HMD2011})
by using corresponding Calibration Database (CALDB v.4.8.5)\footnote{\url{https://cxc.harvard.edu/ciao/download/caldb.html}}.
We first reprocessed the level-2 event files by applying the updated calibration data using CIAO script {\tt chandra\_repro}
\footnote{\url{https://cxc.cfa.harvard.edu/ciao/ahelp/chandra_repro.html}}. After that, we used the CIAO 
tool {\tt axbary} to apply the barycentre correction on the reprocessed level-2 event files. We considered a circular region 
of 2.46 arcsec radius, centred at the source coordinates, to extract the source light curves and spectra. The background 
light curves and spectra were extracted by selecting a circular region of 10 arcsec radius and away from the source. Finally, 
we used {\tt specextract} and {\tt dmextract} to extract the spectra and light curves of the source and background, respectively.
To check the amount of pile-up in each observation, we used the {\tt CIAO} tool {\tt PILEUP\_MAP}. Although it is less
than 10 per cent, we account this effect in spectral fitting by using the convolution model {\tt pileup} \citep{Davis2001}
\footnote{\url{https://heasarc.gsfc.nasa.gov/xanadu/xspec/manual/node304.html}} with all spectral model, with frame time set 
to 0.5s. We did not see any changes with or without {\tt pileup} model.

\label{sec:nustar}
\subsection{NuSTAR}
NGC~6300 was observed with {\it NuSTAR} \citep{Harrison2013} three times; once in 2013, and twice in 2016 (see 
Table~\ref{tab:1} for details). The {\it NuSTAR} consists of two identical modules: FPMA and FPMB. Reduction of 
the raw data was performed with the NuSTAR Data Analysis Software ({\tt NuSTARDAS}, version 1.4.1). Cleaned event 
files were generated and calibrated by using the standard filtering criteria with the {\tt nupipeline} task and 
the latest calibration data files available in the NuSTAR calibration database 
(CALDB)\footnote{\url{http://heasarc.gsfc.nasa.gov/FTP/caldb/data/nustar/fpm/}}. Both the extraction radii for the 
source and the background products were set to be 80 arcsec. With the {\tt nuproduct} task, the spectra and light-curves 
were extracted. The light curves were binned over 100s. Considering the background counts, we limited our spectral 
analysis within $3-40$ keV range. We re-binned the spectra to achieve 20 counts per bin by using the {\tt grppha} task.

\section{Results}
\label{sec:analysis}
We used {\it Suzaku}, {\it Chandra}, and {\it NuSTAR} data between 2007 and 2016 for our analysis. All the 
instruments have different effective area, thus, one needs to take care of this. To address this issue, we used 
`ancillary response file (arf)' in the spectra. For the lightcurves, we used a cross-normalization factor to normalize 
the count rate using crab light curve. We used cross-normalization factor, $N_{FPMA} = 1.0$, $N_{ACIS} = 1.10 \pm 0.05$, 
$N_{XIS} = 0.95 \pm 0.03$ \citep{Madsen15,Madsen17}.

We used the following cosmological parameters in this work: $H_0$ = 70 km s$^{-1}$ Mpc $^{-1}$, $\Lambda_0$ = 0.73, 
and $\Omega_M$ = 0.27 \citep{Bennett2003}.

\subsection{Timing Analysis}

We considered all the X-ray light curves obtained from the {\it Suzaku}, {\it Chandra} and {\it NuSTAR} observations of 
NGC~6300 with 100 s bin for timing analysis. We segregated the total energy bins into various segments for variability 
analysis. We divided the low energy data ($\leq 10$ keV) into two energy bands, namely $0.5-3$ keV and $3-10$ keV for 
variability analysis for {\it Suzaku}/XIS and {\it Chandra} observations. The {\it NuSTAR} data ($3-40$~keV range) were 
divided into two chunks: $3-10$ keV and $10-40$ keV ranges for variability analysis. To examine the time delay between 
the Fe K$_\alpha$ line and continuum, we considered the light curve of $3-40$ keV with the Fe line counts (in 
$6-6.7$~keV energy band) for {\it NuSTAR} observations. The $10-40$~keV light curves were taken into account to analyze 
the variability in the high energy part. 

\begin{table*}
\centering
\caption{Variability statistics based on X-ray observations in different energy ranges 
are shown in this table. In some cases, the average error of observational data exceeds 
the limit of $1\sigma$, resulting negative excess variance. In such cases, we have 
imaginary $F_{var}$, which are not shown in the table.}
\begin{tabular}{lcccccc}
\hline
ID &    $N$ &$x_{max}$ &$x_{min}$&$R$   &$\sigma^2_{NXS}$      &$F_{var}$   \\
                                                                            \\
 & & & & & ($\times10^{-3}$) & \\
\hline
S1(0.5-3)& 198  &0.196  &0.021   &9.40  & -$0.287\pm16.5  $   & $-$ \\
C1(0.5-3)& 18   &0.020  &0.002   &9.11  &  $0.191\pm115.4 $   & $0.17\pm0.36$ \\
C2(0.5-3)& 20   &0.027  &0.005   &6.00   & $0.779\pm53.0  $   & $0.24\pm0.15$ \\
C3(0.5-3)& 19   &0.022  &0.002   &9.88   & $1.180\pm73.3  $   & $0.36\pm0.16$ \\
C4(0.5-3)& 21   &0.028  &0.002   &12.1   & $1.510\pm55.8  $   & $0.35\pm0.13$ \\
C5(0.5-3)& 20   &0.020  &0.002   &9.11   &-$0.361\pm59.7  $   & $-$ \\
&&&&&&\\
S1(3-10)& 203  &0.49  &0.11   &4.31   & $7.41\pm3.73$  & $0.19\pm0.02$ \\
C1(3-10)& 21   &0.09  &0.02   &4.55   & $2.54\pm10.3$  & $0.19\pm0.06$ \\
C2(3-10)& 20   &0.30  &0.15   &2.08   & $3.47\pm3.18$  & $0.13\pm0.03$ \\
C3(3-10)& 22   &0.21  &0.01   &2.32   & $12.9\pm5.36$  & $0.32\pm0.05$ \\
C4(3-10)& 21   &0.30  &0.12   &2.56   & $66.1\pm3.41$  & $0.18\pm0.04$ \\
C5(3-10)& 21   &0.25  &0.07   &3.65   & $8.33\pm3.83$  & $0.22\pm0.04$ \\
N1(3-10)& 44   &1.01  &0.32   &3.17   & $35.9\pm1.89$  & $0.25\pm0.03$ \\
N2(3-10)& 49   &0.74  &0.20   &3.75   & $24.9\pm1.78$  & $0.23\pm0.03$ \\
N3(3-10)& 62   &0.84  &0.31   &2.68   & $14.7\pm4.98$  & $0.16\pm0.03$ \\
&&&&&&\\
N1(10-40)& 44   &0.63   &0.27   &2.33 & $13.1\pm2.67$  & $0.18\pm0.03$ \\
N2(10-40)& 49   &0.52   &0.19   &2.61 & $11.5\pm2.23$  & $0.19\pm0.02$ \\
N3(10-40)& 62   &0.93   &0.26   &3.60 &-$1.25\pm12.0$  & $-$ \\
&&&&&&\\
\hline
\end{tabular}
\label{tab:fvar}
\end{table*}

\subsubsection{\bf Variablility}
To examine the temporal variability in X-ray emission from NGC~6300, in the different energy bands during the period 
of 17th October 2007 to 24th August 2016, we estimated numerous parameters. The fractional variability $F_{var}$ 
(\citet{Edelson1996}; \citet{Nandra1997}; \citet{Edelson2001}; \citet{Edelson2012}; \citet{Vaughan2003}; \citet{RP1997}) 
for a light curve of $x_i$ count/s with finite measurement error $\sigma_i$ of length $N$ with a mean $\mu$ and standard
deviation $\sigma$ is given by:

\begin{equation}
F_{var}=\sqrt{\frac{\sigma^2_{XS}}{\mu^2}}
\end{equation}
where $\sigma^2_{XS}$ is excess variance (\citet{Nandra1997}; \citet{Edelson2002}), an estimator of 
the intrinsic source variance and is given by:
\begin{equation}
\sigma^2_{XS}=\sigma^2 - \frac{1}{N}\sum_{i=1}^{N} \sigma^2_{i}.
\end{equation} 

The normalized excess variance is given by $\sigma^2_{NXS}=\sigma^2_{XS}/\mu^2$. The uncertainties in $\sigma^2_{NXS}$ 
and $F_{var}$ are taken from \citet{Vaughan2003} and \citet{Edelson2012}. The peak to peak amplitude is defined as 
$R=x_{max}/x_{min}$ (where $x_{max}$ and $x_{min}$ are the maximum and minimum flux, respectively) to investigate 
the variability in the X-ray light curves. 

The X-ray photons from NGC~6300 in different energy bands (0.5--3 keV, 3--10 keV, 10--40 keV ranges) showed 
different magnitude of variabilities. The results are shown in Table~\ref{tab:fvar}. From the low energy part 
($0.5-3$ keV range), we obtained an average $R$ value of $9.27$ which has a range from $6.00$ to $12.1$. 
However, for higher energy part (3--10 keV range), we observed $<R>=3.02$, ranging from $2.08$ to $4.55$. 
Thus, the high energy part showed fewer variabilities in terms of the average values and range of $R$. 
Although, $\sigma^2_{NXS}$ exhibited opposite nature. A trend of increasing normalized excess variance 
can be seen from Table~\ref{tab:fvar}. It should also be noted that the errors of $\sigma^2_{NXS}$ (calculated
considering variance $\sim \overline{\sigma^2_{err}}$) are larger than the values for 0.5--3 keV energy band. 
Thus, the conclusions solely based on $\sigma^2_{NXS}$ would be erroneous.

We calculated the fractional variability ($F_{var}$) in terms of normalized excess variance ($\sigma^2_{NXS}$) 
to investigate the variabilities in different wavebands. These quantities describe the variability strength present
in AGN light curves. Though $\sigma^2_{NXS}$ and $F_{var}$ represent similar information, the latter was
used as $F_{var}$ is independent of the signal-to-noise ratio of the light curve. From Table \ref{tab:fvar}, we 
can infer that NGC~6300 had a constant decrease of variability strength with increasing energy ($<F_{var}^{0.5-3}>
=0.28$; $<F_{var}^{3-10}>=0.208$; $<F_{var}^{10-40}>=0.185$) over the entire period of time.

\subsubsection{\bf Correlation}
In order to investigate the physical connection between the Fe line and the X-ray continuum, we computed the Pearson
correlation coefficient ($r$) and Spearman's rank correlation coefficient ($r_s$) using the {\it NuSTAR}
observation. Quantitive analysis of the correlation based on the Pearson coefficient provided a degree of linear 
correlation between the Fe-line emission and the X-ray continuum. We correlated $3-40$~keV light curves 
(for X-ray continuum) with the Fe line light curves. The Fe line light curves are computed in the energy range of 
$6-6.7$~keV. The positive and negative values of the Pearson coefficient for all three observations indicated a 
very weak linear correlation between the Fe-flux and the X-ray continuum. This implies a plausible disjoint mechanism 
involved in the emission of the Fe line and the X-ray continuum. We estimated the Spearman's rank correlation $(r_s)$ 
to further quantify the degree of correlation between the Fe flux and X-ray continuum. It reflects a similar result 
for the first two observations. The result of this correction study is quoted in Table~\ref{tab:correlation}.

\begin{figure}
\includegraphics[height=1.0\columnwidth,width=1.3\columnwidth]{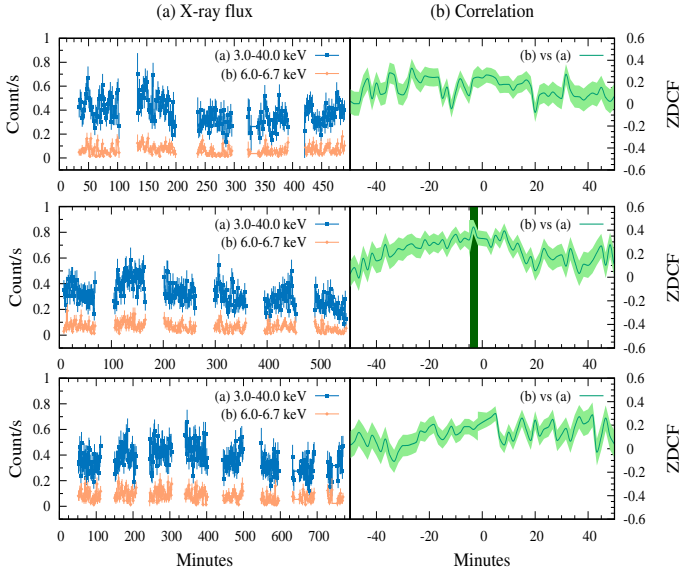}
\caption{The ZDCF analysis curves showing the correlation as a function of time-delay 
between the X-ray light curves and the Fe-line flux.}
\centering
\label{zdcf}
\end{figure}

We also investigate the time-delay between the Fe-line flux and the X-ray continuum by using the 
$\zeta$-transformed discrete correlation function (ZDCF) method fully described by \citet{Alexander1997}
\footnote{\burl{http://www.weizmann.ac.il/particle/tal/research-activities/software}}. The ZDCF-code is
publicly available for estimating the cross-correlation function of unevenly sampled light curves. We 
considered the X-ray light curves in $3-40$ keV range with the Fe-line flux for three NuSTAR observations 
to estimate the ZDCF coefficient. We calculated ZDCF for two different cases: omit zero lag points and include
zero lag points, and in both cases, we got a similar result. The values of ZDCF coefficient with time delay
is presented in Table \ref{tab:correlation}. Also, the variation of coefficient with time delay is shown in 
Figure~\ref{zdcf}. Although there is no prominent peak in the correlation function, we found a moderate peak 
at the time delay of $1.67\pm1.5$ minutes for the first observation with ZDCF coefficient value of $0.18 \pm 0.12$
and peak at $-3.30 \pm 1.5$ minutes for second observation with ZDCF coefficient value of $0.45\pm 0.12$. 
In the case of third observation, we did not notice any peak in the correlation function.

\begin{table}
\centering
\caption{Pearson correlation coefficient ($r$), Spearman's rank correlation coefficient ($r_s$) between the 
Fe-line flux and X-ray light curves. Corresponding probabilities of the null-hypothesis ($p$-value) are computed
for N-2 degrees of freedom (DOF). The last two columns show the results of the ZDCF analysis between Fe-line flux
and X-ray light curves.}

\label{tab:correlation}
\begin{tabular}{lccccc}
\hline
ID   &    $r$   &  $r_s$  &  $p-value$     &    $ZDCF$         &     $time-delay$ \\
 & & & & & (min) \\
\hline
N1   &  0.2385  &  0.1835 &    .0115       &  $0.265\pm0.122$  &   $1.67\pm1.5$    \\
N2   &  0.3490  &  0.3494 &  < .001        & $0.445\pm0.123$   &   $-3.30\pm1.5$    \\
N3   & -0.2700  &  0.3676 &  < .001        &      $-$          &   $-$  \\
\hline
\end{tabular}
\end{table}

\subsection{Spectral Analysis}

Spectral analysis of data obtained from the {\it Suzaku}, {\it Chandra} and {\it NuSTAR} observations of
NGC~6300 was carried out by using the software package {\tt XSPEC} v12.10 \citep{Arnaud1996}. For spectral 
fitting, we explored several phenomenological and physical models in order to understand the core region 
of NGC~6300. We used {\tt powerlaw}, {\tt compTT} \citep{Titarchuk1994}, {\tt pexrav} \citep{MZ95}, 
and {\tt MYTORUS} \citep{MY2009} models to approximate the primary continuum and reflection components. 
While fitting the data with the first three continuum models separately, a {\tt Gaussian} component is considered 
for the iron fluorescent emission line. The soft X-ray excess observed in the spectra of many AGNs, is usually 
modelled with the combination of {\tt powerlaw} and {\tt APEC} model. The normalization parameter was tied with 
the primary component in order to get an idea of scattering fraction $f_s$; modelled with `{\tt constant}' in 
{\tt XSPEC}. The soft excess component in {\tt XSPEC} reads as: {\tt constant*(powerlaw + apec)}. Along with 
these components, we used two absorption components, namely {\tt TBabs} and {\tt zTBabs} \citep{Wilms2000}. 
{\tt TBabs} was used for Galactic absorption and hydrogen column density ($N_{H,Gal}$) was fixed at 
$8.01 \times 10^{20}$ $cm^{-2}$ \citep{HI4PI2016}. We calculated error for each spectral parameters 
with 90 per cent confidence level (1.6 $\sigma$). The errors are calculated using `{\tt error}' command in {\tt XSPEC}.

\label{sec:powerlaw}
\subsubsection{{\bf Power law}}
We started the spectral analysis with the simple {\tt powerlaw} model. Our baseline model in {\tt XSPEC} 
reads as:

{\tt TBabs1*(zTBabs2*(zpowerlaw + zGaussian) + soft excess)}.

\begin{figure*}
\includegraphics[width=\linewidth]{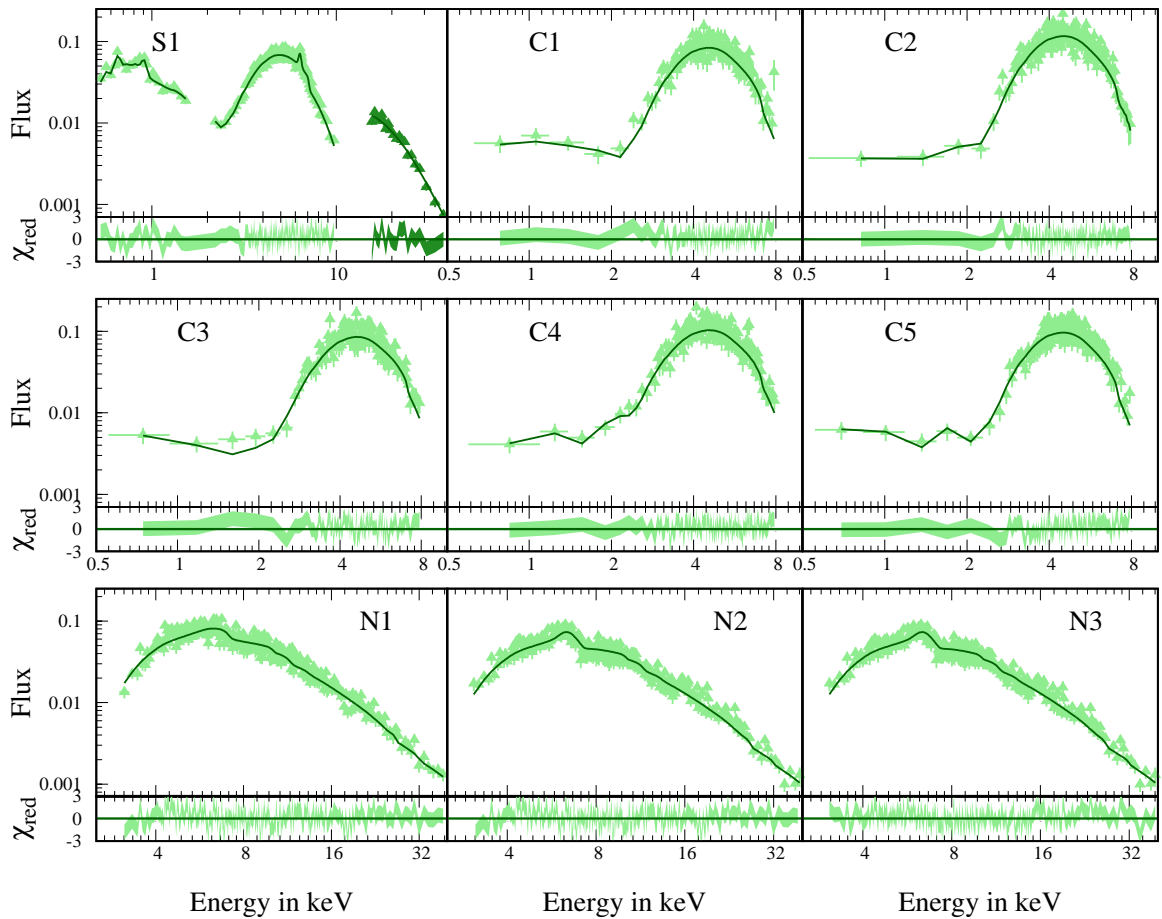}
\caption{Power law model fitted spectra (upper panels) of NGC~6300 from the {\it Suzaku}, {\it Chandra} 
and {\it NuSTAR} observations along with the residuals obtained from the spectral fitting (bottom panels).}
\label{fig:pl-spec}
\end{figure*}

We started our analysis using data from the {\it Suzaku} observation in 2007. The $0.5-40$~keV spectrum was 
fitted with the above spectral model. The parameters obtained from the fitting are $N_H = 2.09 \times 10^{23}$ 
$cm^{-2}$, $\Gamma = 1.81$, an Fe $K\alpha$ line at $6.37$~keV with equivalent width (EW) of 114 eV and the 
reduced chi-square ($\chi^2/dof$) = 1.03 (for 2549 dof). Next, we analyzed the data from the 2009 epoch when
NGC~6300 was observed five times within 11 days with the {\it Chandra} observatory. We did not detect Fe K$\alpha$ 
line in all five spectra. We verified this by using the {\tt ftest} task in {\tt XSPEC}. We fitted all the {\it Chandra} 
spectra by removing the {\tt Gaussian} component from the baseline model. The hydrogen column density ($N_H$) along 
the line of sight varied between $1.77 \times 10^{23}$ $cm^{-2}$ and $1.89 \times 10^{23}$ $cm^{-2}$. The power-law 
photon indices were similar during the 2009 epoch of observations. NGC~6300 was observed with the {\it NuSTAR} 
observatory in February 2013. The data obtained from the {\it NuSTAR} observation were fitted with the model 
yielding parameters such as $N_H = 1.49 \times 10^{23}$ $cm^{-2}$, $\Gamma = 1.61$, an Fe $K\alpha$ line at $6.36$~keV 
with an equivalent width of 258 eV and the $\chi^2/dof$ = 758/702. The iron line was found to be broader than the 
previous observations. The spectra became harder with photon indices of $\Gamma = 1.52$ and $1.54$ during the 
observations in Jan 2016 and Aug 2016, respectively. The iron line width was increased to $280$ and $299$~eV during 
these two observations. However, the hydrogen column density was decreased in the 2016 epochs to $1.08 \times 10^{23}$ 
$cm^{-2}$ which is almost $\sim 50$ per cent of the value of the column density during 2007 epoch of observation. In 
Figure~\ref{fig:pl-spec}, the {\tt powerlaw} model fitted spectra are shown. In Figure~\ref{fig:contour}, the contour 
plots for $\Gamma$ vs $N_H$ are shown for S1, C2, \& N2. The {\tt powerlaw} model fitted spectral analysis result is 
shown in Table~\ref{tab:PL}.

\label{sec:comptt}
\subsubsection{{\bf CompTT model}}

The powerlaw model provided valuable information on the variations of spectral hardness and hydrogen column
density over the observation duration of $\sim 9$ years. However, the fundamental properties such as electron
temperature ($kT_e$), optical depth of the medium ($\tau$), and approximate shape of accretion geometry (slab
or spherical) are necessary to gain a deeper understanding of the system. To estimate these quantities, we 
replaced the {\tt powerlaw} model with a more physical {\tt compTT} model \citep{Titarchuk1994} in our spectral
fitting. The X-ray emitting Compton cloud is characterized by the hot electron temperature ($kT_e$) and optical 
depth ($\tau$). The model considered here can be expressed as:

{\tt TBabs1*(zTBabs2 * (compTT + zGaussian) + soft excess)}.

While fitting with the {\tt compTT} model, the $\chi^2$ values obtained were acceptable for all observations. We 
found that the electron temperature varied over the years, indicating a change in the spectral state. In fact, 
$kT_e$ increased over the period of $\sim 9$ years from $37.2$~keV to $88.6$~keV. The value of the optical depth 
varied dramatically over the entire span of the observation duration. The optical depth remained mostly constant
during the 2009 epoch except for the {\it Chandra} observation on 3 June 2009 (`C1'). Particularly on that date,
we found significantly high electron temperature (much higher than the {\it Suzaku} observation - S1) along with 
the higher value of the optical depth, suggesting a denser and hotter Compton cloud around the black hole. 

Results presented here were obtained by considering a spherical cloud with seed photon temperature fixed at $30$~eV,
which is likely to be the inner disc temperature for a BH with a mass of $\sim 10^7$ $M_{\odot}$. We also considered
the slab geometry for which the parameter variations remained reasonably consistent as obtained with the spherical geometry. 
Thus, the geometry of the Compton cloud remained unclear with the {\tt compTT} model.

\label{sec:pexrav}
\subsubsection{{\bf Pexrav}}

The reported inclination for of NGC~6300 is around $77^{\circ}$ \citep{Leighly99}. As a high inclination source,
the X-ray spectrum from the inner region is more prone to suffer reflection from the disc. To estimate the reflection 
coefficient over the entire period of our observation, we applied reflection model {\tt pexrav} \citep{MZ95}. The 
{\tt pexrav} model contains a powerlaw continuum and a reflected component from an infinite neutral slab. We estimated 
relative reflection ($R$) of the source. We fixed the photon index ($\Gamma_{pexrav}$) with the value of $\Gamma$ obtained 
from the {\tt powerlaw} model. We fixed abundances for heavy elements and iron at Solar value (i.e. 1). In the beginning, 
we froze $\cos i$ at 0.22 ($\theta_{obs} = 77$\textdegree). Later, we re-analyzed the data by fixing $\cos i$ at 0.15 
($\theta_{obs} = 80$\textdegree; obtained from {\tt MYTROUS} model fit, see~\ref{sec:MYT}). As we could not constrain 
the cut-off energy of the Compton cloud, we froze it at $1000$~keV. We found that during the 2007 {\it Suzaku} observation, 
the reflection was moderate ($R=0.62$). However, in 2009 epoch, $R$ increased and dominated during the three observations
($R>1$). During the other two observations in 2009 epoch,  $R$ is estimated to be in the range of 0.8 and 1,
indicating a strong reflection. The reflection hump was seen in $\sim 15-30$~keV energy range; thus, the value of $R$ 
obtained is in the range of 0.5 to 8.0 from the {\it Chandra} data and are not well constrained. The $3-40$~keV spectra 
from the {\it NuSTAR}observations in 2013 \& 2016 epochs gave us a good estimation of $R$. During all three {\it NuSTAR}
observations, we found that the reflection was strong with $R>0.74$. The model fitted results are given in Table~\ref{tab:PL}.

\begin{figure*}
\includegraphics[width=\linewidth]{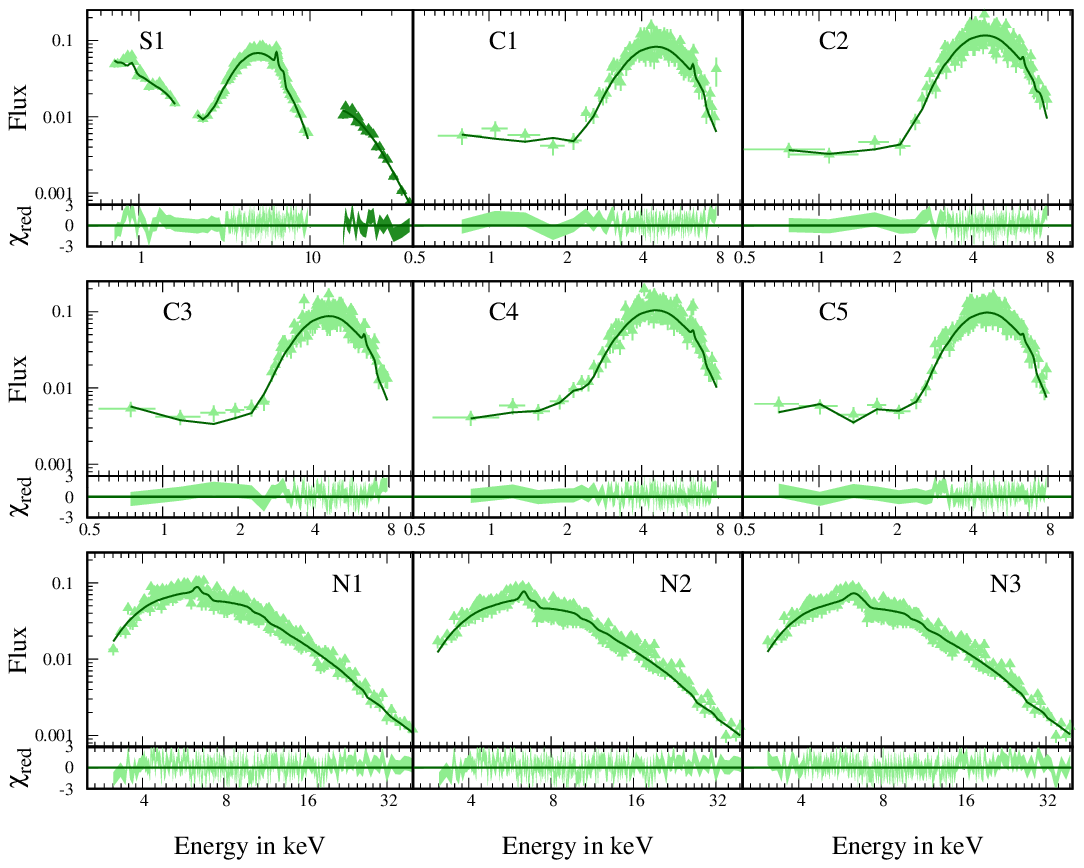}
\caption{My-Torus decoupled model fitted spectra (upper panels) of NGC~6300 from the {\it Suzaku}, {\it Chandra} 
and {\it NuSTAR} observations along with the residuals obtained from the spectral fitting (bottom panels).}
\label{fig:mytd-spec}
\end{figure*}

\label{sec:MYT}

\subsubsection{{\bf MYTORUS}}

The AGNs are surrounded by circumnuclear absorbing clouds popularly referred to as `torus'. 
The phenomenological models do not consider the complex structure of the `torus'. Several 
physical models have been developed which account for the `toroidal geometry' and calculate 
reflection and line spectra self-consistently \citep{Ikeda2009,MY2009,Brightman2011,Balokovic2018}. 
We studied NGC~6300 with a physical model, namely the {\tt MYTORUS}\footnote{\url{https://www.mytorus.com/}} 
model \citep{MY2009,Yaqoob2012} which describes an absorbing torus surrounding the nuclear region 
with half opening angle fixed at 60\textdegree. This model consists of three components: absorbed 
primary continuum or {\it zeroth-ordered} component ({\tt MYTZ}), a scattered/reflected component 
({\tt MYTS}) and an iron fluorescent line component ({\tt MYTL}; Fe K$\alpha$ and Fe K$\beta$ line). 
The {\tt MYTORUS} model can be used in two configurations: {\tt coupled} and {\tt decoupled} \citep{Yaqoob2012}. 
The {\tt coupled} configuration describes a uniform circumnuclear torus while the {\tt decoupled} 
configuration attributes the nonuniformity in the torus shape and density profiles.

In the beginning, we tested the default or {\tt coupled} configuration of the {\tt MYTORUS} model. 
The model reads as,

{\tt TBabs(powerlaw*MYTZ + $A_S$MYTS + $A_L$MYTL + soft excess)}.

In this configuration, the equatorial hydrogen column density ($N_{H,eq}$), photon index of the 
incident primary continuum ($\Gamma$), inclination angle ($\theta_{obs}$), and model normalization 
are tied together. As recommended, the relative normalizations of the scattered and line components 
are tied together, i.e. $A_S = A_L$. We allowed the inclination angle ($\theta_{obs}$) to vary freely. 
We found that the inclination angle varied between 76\textdegree and 82\textdegree. With this model, 
we achieved a good $\chi^2$ fit for all observations. The photon indices and equatorial column density 
obtained using this model followed a similar trend as obtained from the {\tt powerlaw} model. In the 
{\tt MYTROUS} model, the equivalent width is not a free parameter; instead, it is computed self-consistently. 
The full width half maxima (FWHM) and flux ($F_{K\alpha}$) of iron K$\alpha$ line are obtained 
from a Gaussian convolution model {\tt gsmooth} which can be used with {\tt MYTL} component (for details, see
\citet{Yaqoob2012}). The results obtained using this model are presented in Table~\ref{tab:mytc}.

Later, we tested the {\tt decoupled} configuration of the {\tt MYTORUS} model. This can be achieved 
by decoupling the column density of different components. Initially, we fixed $\theta$ of {\tt MYTZ} 
component to 90\textdegree. This component describes the absorbed transmitted component. The hydrogen 
column density of {\tt MYTZ} will provide us with a column density ($N_{H,Z}$) along the line-of-sight. 
Next, we fixed inclination angle of {\tt MYTS} and {\tt MYTL} at 0\textdegree. The column density is 
decoupled from the {\tt MYTZ} component. The column density of scattered component describes averaged
global column density ($N_{H,S}$). This component describes a component which is scattered from the 
backside and coming to us through the hole of a patchy torus (see Figure~2 of \cite{Yaqoob2012}). 
The model can be read as:

{\tt TBabs(powerlaw * MYTZ + $A_{S00}$MYTS + $A_{L00}$MYTL+ soft excess}).

\begin{figure*}
\begin{subfigure}[t]{0.33\textwidth}
    \includegraphics[width=4.5cm,angle=0]{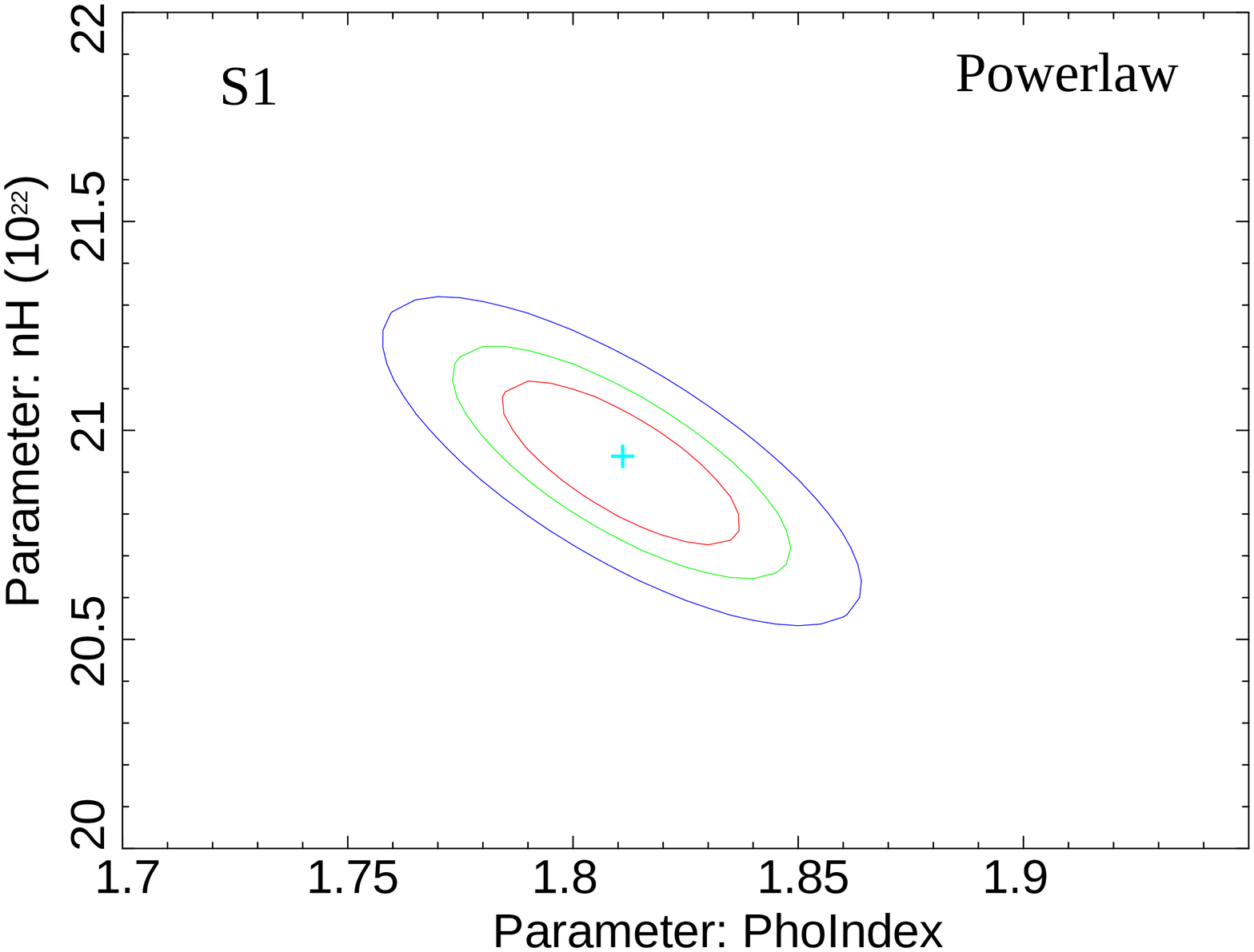}
\label{fig:figure14_1}
\end{subfigure}\hfill
\begin{subfigure}[t]{0.33\textwidth}
  \includegraphics[width=4.5cm,angle=0]{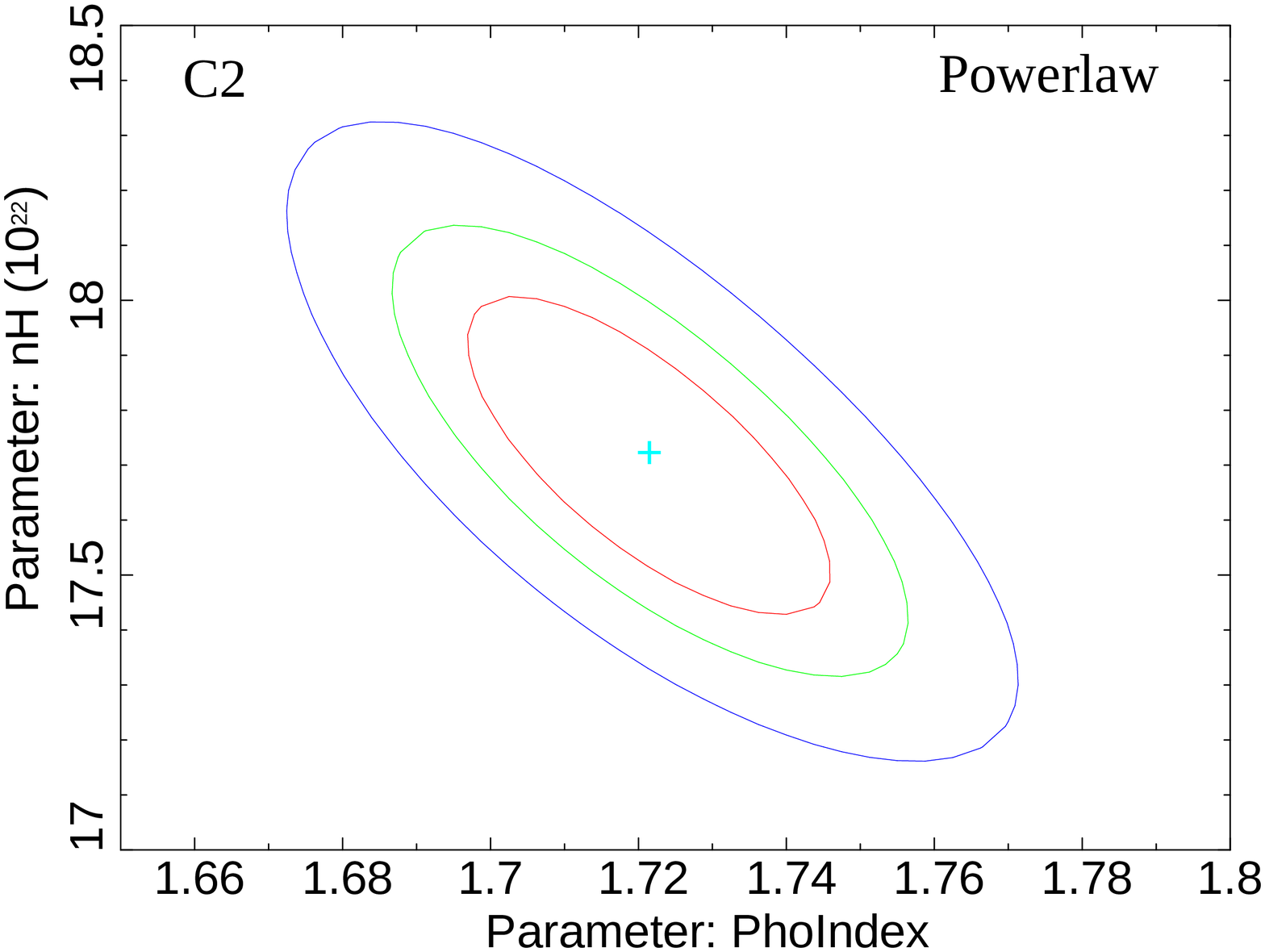}
\label{fig:figure14_2}
\end{subfigure}\hfill
\begin{subfigure}[t]{0.33\textwidth}
    \includegraphics[width=4.5cm,angle=0]{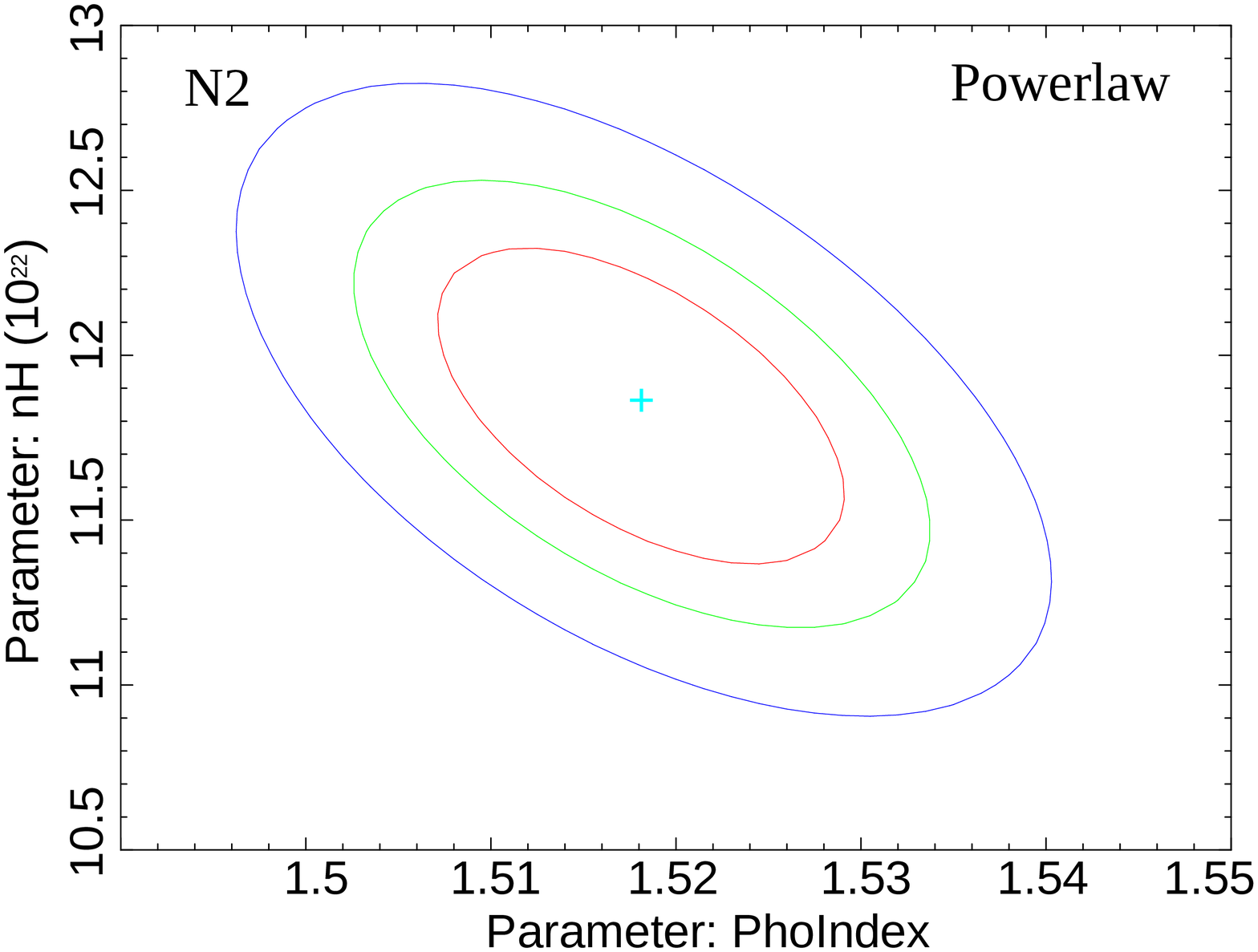}
\label{fig:figure14_3}
\end{subfigure}

\begin{subfigure}[t]{0.33\textwidth}
    \includegraphics[width=4.5cm,angle=0]{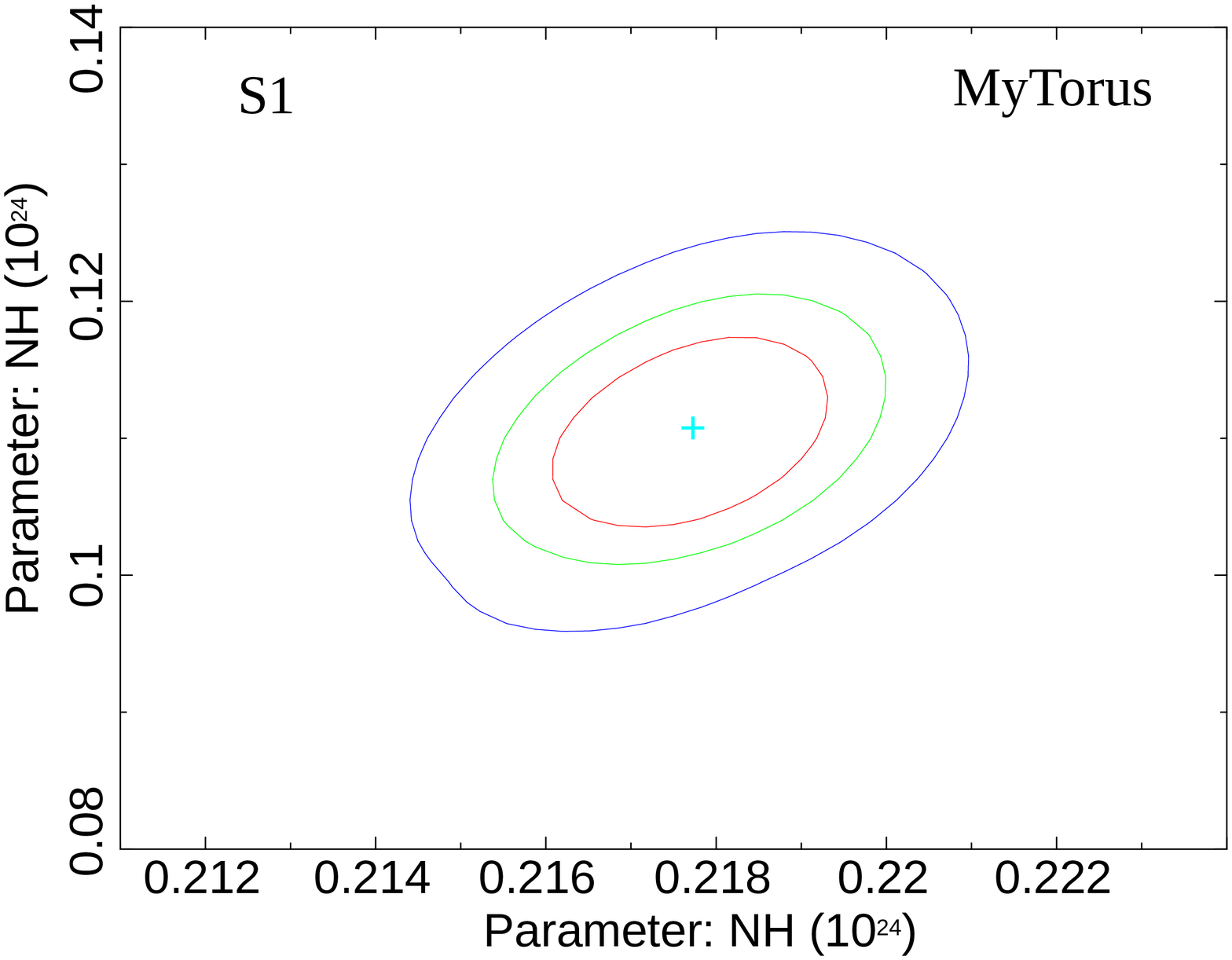}
\label{fig:figure14_4}
\end{subfigure}\hfill
\begin{subfigure}[t]{0.33\textwidth}
    \includegraphics[width=4.5cm,angle=0]{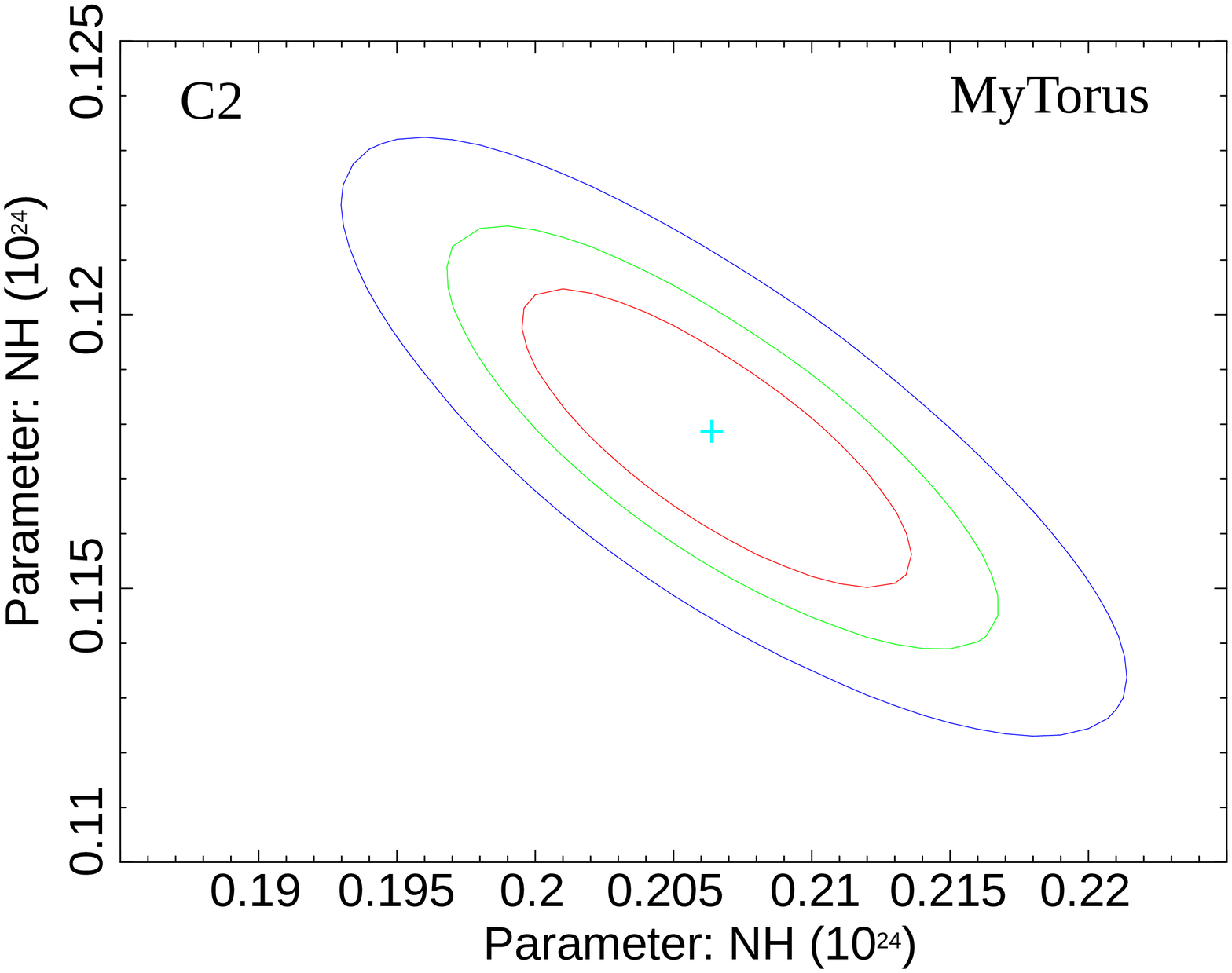}
\label{fig:figure14_5}
\end{subfigure}\hfill
\begin{subfigure}[t]{0.33\textwidth}
    \includegraphics[width=4.5cm,angle=0]{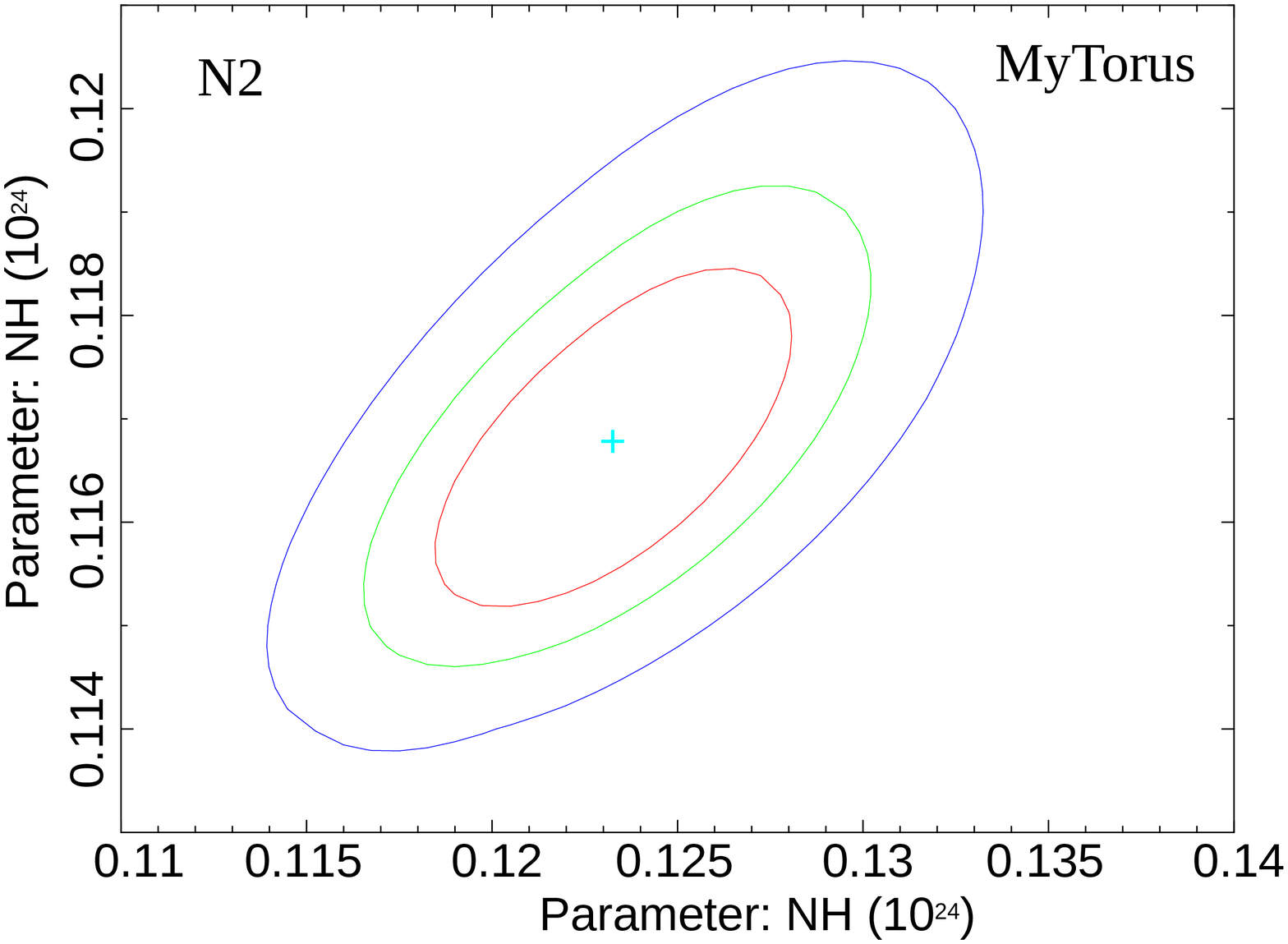}
\label{fig:figure14_6}
\end{subfigure}

\caption{In the upper panel, contour plots are shown between photon index ($\Gamma$) and line-of-sight
hydrogen column density ($N_H$), obtained from the spectral analysis with {\tt powerlaw} model, for S1,
C2, and N2. In the lower panel, contour plots are shown between line-of-sight column density ($N_{H,z}$)
and global-averaged column density ($N_{H,S}$), obtained from the spectral analysis with the 
decoupled-{\tt MYTORUS} model for S1, C2, and N2. The solid red, green and blue lines correspond to 68, 
90 and 99 per cent confidence levels, respectively.}
 \label{fig:contour}
\end{figure*}

The {\tt Decoupled} configuration did not give us a significantly deviated fit from the {\tt Coupled} model. 
$\Gamma$ was found to follow the same trend as the {\tt Coupled} model. The line-of-sight column density 
($N_{H,Z}$) was observed to follow the same pattern as $N_{H,eq}$. However, the averaged-global column density 
($N_{H,S}$) was roughly constant for all nine observations. The scattering fraction ($A_S$) was 1.06 in 2007, 
which indicates that the reflection was dominating and delayed. In 2009 epoch, $A_S < 1$, which infers a weaker 
reflection compared to the primary emission. In 2013, \& 2016 epochs, we observed $A_S >2$ from this model. 
This indicates that the spectra were dominated by the reflection. We calculated FWHM and line flux for
Fe K$\alpha$ line emission from the Gaussian convolution model {\tt gsmooth}. We also calculated intrinsic 
luminosity ($L_{int}$) of the source using `{\tt clum}' command in {\tt XSPEC} in the energy band of $2-10$~keV.
We computed the intrinsic luminosity for {\tt MYTZ} component \citep{Yaqoob2012}. The results of this model are 
tabulated in Table~\ref{tab:mytd}. The {\tt MYTORUS} model fitted ({\tt decoupled} configuration) spectra 
are shown in Figure~\ref{fig:mytd-spec}. In Figure~\ref{fig:contour}, we show the contour plot for $N_{H,Z}$ vs 
$N_{H,S}$.

% \newpage

\begin{landscape}
\begin{table}
\caption{Phenomenological model fitted results}
\label{tab:PL}
\begin{tabular}{lcccccccccccccc}
\hline
\hline
ID& $N_H$ & $\Gamma$ & PL. norm & Line E& $F_{K\alpha}$& EW & $f_s$ &$kT^{apec}$ & $\chi^2$/dof&$kT_e^{compTT}$ & $\tau$ & $R$ & $F_{2-10 keV, obs}$ \\
 &($10^{22}$ cm$^{-2}$)& &($10^{-3}$ ph cm$^{-2}$ s$^{-1}$)&(keV) & ($10^{-5}$ ph cm$^2$ s$^{-1}$)&(eV) &($10^{-2}$) &(keV)& &(keV)&  & & ($10^{-11}$ ergs cm$^{-2}$ s$^{-1}$) \\
\hline
S1 & $ 20.9^{+0.3}_{-0.3}$ & $ 1.81^{+0.04}_{-0.03} $ & $ 33.2^{+0.24}_{-0.22}$  & $ 6.37^{+0.08}_{-0.12} $ & $ 1.52^{+0.09}_{-0.10} $ & 114 & $ 1.94^{+0.17}_{-0.13}$ & $ 0.23^{+0.06}_{-0.08} $       & 2614/2549 &$ 37.2^{+2.1}_{-1.8} $&$ 3.08^{+0.22}_{-0.25} $ &$ 0.62^{+0.10}_{-0.09} $ &$ 1.68^{+0.04}_{-0.14}$ \\
C1 & $ 18.9^{+0.3}_{-0.3}$ & $ 1.69^{+0.05}_{-0.05} $ & $ 6.94^{+0.33}_{-0.30}$  & $-$ & $-$ &  $-$ & $ 1.68^{+0.12}_{-0.17}$ & $ 0.35^{+0.07}_{-0.11} $       &  283/ 293 &$ 44.7^{+1.8}_{-2.3} $&$ 2.97^{+0.19}_{-0.34} $ &$ 1.09^{+0.13}_{-0.14} $ &$ 1.23^{+0.22}_{-0.02}$ \\
C2 & $ 17.7^{+0.4}_{-0.4}$ & $ 1.72^{+0.04}_{-0.04} $ & $ 8.06^{+0.64}_{-0.79}$  & $-$ & $-$ &  $-$ & $ 2.58^{+0.14}_{-0.10}$ & $ 0.45^{+0.10}_{-0.07} $       &  305/ 321 &$ 46.6^{+3.1}_{-4.1} $&$ 0.39^{+0.05}_{-0.09} $ &$ 1.35^{+0.22}_{-0.19} $ &$ 1.68^{+0.35}_{-0.21}$ \\
C3 & $ 18.3^{+0.7}_{-0.6}$ & $ 1.78^{+0.06}_{-0.05} $ & $ 8.04^{+0.14}_{-0.13}$  & $-$ & $-$ &  $-$ & $ 2.35^{+0.08}_{-0.12}$ & $ 0.44^{+0.10}_{-0.13} $       &  289/ 293 &$ 43.1^{+5.4}_{-5.4} $&$ 0.51^{+0.07}_{-0.10} $ &$ 0.83^{+0.11}_{-0.15} $ &$ 1.52^{+0.33}_{-0.21}$ \\
C4 & $ 18.5^{+0.6}_{-0.5}$ & $ 1.78^{+0.06}_{-0.04} $ & $ 7.72^{+0.47}_{-0.37}$  & $-$ & $-$ & $-$ & $ 1.78^{+0.16}_{-0.19}$ & $ 0.37^{+0.07}_{-0.12} $       &  295/ 315 &$ 44.1^{+5.1}_{-4.9} $&$ 0.49^{+0.09}_{-0.05} $ &$ 1.74^{+0.19}_{-0.22} $ &$ 1.62^{+0.24}_{-0.10}$ \\
C5 & $ 18.3^{+0.5}_{-0.3}$ & $ 1.79^{+0.04}_{-0.04} $ & $ 9.38^{+0.42}_{-0.49}$  & $-$ & $-$ &  $-$ & $ 1.61^{+0.22}_{-0.27}$ & $ 0.47^{+0.05}_{-0.05} $       &  306/ 299 &$ 46.1^{+6.2}_{-5.4} $&$ 0.89^{+0.14}_{-0.14} $ &$ 0.82^{+0.07}_{-0.10} $ &$ 1.44^{+0.16}_{-0.25}$ \\
N1 & $ 14.9^{+0.3}_{-0.3}$ & $ 1.61^{+0.02}_{-0.02} $ & $ 4.67^{+0.41}_{-0.53}$  & $ 6.36^{+0.29}_{-0.34} $ & $ 1.18^{+0.30}_{-0.24} $ & 258 & $     ^{     }-{     }$ & $     ^{     }-{     } $       &  758/ 702 &$ 55.6^{+1.9}_{-2.2} $&$ 0.77^{+0.09}_{-0.08} $ &$ 0.86^{+0.13}_{-0.17} $ &$ 1.59^{+0.03}_{-0.05}$ \\
N2 & $ 11.7^{+0.8}_{-0.5}$ & $ 1.52^{+0.02}_{-0.02} $ & $ 3.46^{+0.25}_{-0.33}$  & $ 6.38^{+0.10}_{-0.09} $ & $ 7.39^{+0.11}_{-0.17} $ & 280 & $     ^{     }-{     }$ & $     ^{     }-{     } $       &  802/ 715 &$ 79.6^{+3.2}_{-2.8} $&$ 1.12^{+0.09}_{-0.16} $ &$ 1.05^{+0.16}_{-0.19} $ &$ 1.27^{+0.04}_{-0.02}$ \\
N3 & $ 10.7^{+0.5}_{-0.6}$ & $ 1.54^{+0.04}_{-0.04} $ & $ 4.59^{+0.37}_{-0.44}$  & $ 6.32^{+0.13}_{-0.15} $ & $ 9.29^{+0.14}_{-0.10} $ & 299 & $     ^{     }-{     }$ & $     ^{     }-{     } $       &  781/ 821 &$ 88.6^{+2.8}_{-3.5} $&$ 0.54^{+0.06}_{-0.11} $ &$ 0.74^{+0.06}_{-0.20} $ &$ 1.60^{+0.02}_{-0.06}$ \\
\hline
\end{tabular}
\end{table}
% \end{landscape}

% \begin{landscape}
\begin{table}
\caption{{\tt MYTORUS} model fitted results -- {\tt Coupled} configuration}
\label{tab:mytc}
\begin{tabular}{lcccccccccccc}
\hline
\hline
ID &$\Gamma $&PL. norm &$N_{H,eq}$ &$\theta_{obs}$ & $A_S = A_L $ & EW &$F_{K\alpha}$ &FWHM &$f_s$& $kT^{apec}$  &$\chi^2$/dof \\
   & &($10^{-3}$ ph cm$^{-2}$ s$^{-1}$)&($10^{24}$ cm$^{-2}$)&(degree)& & (eV) &($10^{-13}$ ergs cm$^{-2}$ s$^{-1}$) &(km s$^{-1}$)&($10^{-2}$) &(keV)&  \\
\hline
S1& $ 1.84^{+0.04}_{-0.04} $ & $ 38.2^{+3.42}_{-2.92} $ & $ 0.232^{+0.009}_{-0.010}$ & $ 78.91^{+2.67}_{-2.75}$ & $ 1.36^{+0.08}_{-0.03}$ & $119^{+13}_{-10} $      &$ 1.81^{+0.13}_{-0.14} $ & $  5290^{+529 }_{-471 }$ & $ 1.30^{+0.14}_{-0.12}$ & $ 0.24^{+0.02}_{-0.01}$ &2476/2380 \\ 
C1& $ 1.73^{+0.05}_{-0.03} $ & $ 8.08^{+0.43}_{-0.41} $ & $ 0.207^{+0.009}_{-0.006}$ & $ 79.20^{+2.56}_{-2.41}$ & $ 1.15^{+0.10}_{-0.13}$ & $112^{+9 }_{-12} $      &$ 1.94^{+0.12}_{-0.10} $ & $    71^{+5   }_{-7   }$ & $ 1.35^{+0.10}_{-0.15}$ & $ 0.29^{+0.03}_{-0.04}$ & 278/292  \\
C2& $ 1.75^{+0.05}_{-0.06} $ & $ 11.6^{+0.41}_{-0.49} $ & $ 0.199^{+0.007}_{-0.012}$ & $ 79.98^{+2.09}_{-1.94}$ & $ 0.62^{+0.05}_{-0.10}$ & $ 96^{+8 }_{-10} $      &$ 1.43^{+0.10}_{-0.08} $ & $    52^{+6   }_{-3   }$ & $ 4.48^{+0.17}_{-0.21}$ & $ 0.28^{+0.04}_{-0.03}$ & 311/314  \\
C3& $ 1.76^{+0.03}_{-0.04} $ & $ 8.86^{+0.25}_{-0.32} $ & $ 0.196^{+0.004}_{-0.008}$ & $ 76.17^{+2.92}_{-3.04}$ & $ 1.00^{+0.16}_{-0.22}$ & $ 88^{+10}_{-7 } $      &$ 1.93^{+0.15}_{-0.19} $ & $    45^{+8   }_{-4   }$ & $ 1.25^{+0.25}_{-0.28}$ & $ 0.37^{+0.07}_{-0.04}$ & 295/294  \\
C4& $ 1.79^{+0.09}_{-0.07} $ & $ 9.28^{+0.14}_{-0.16} $ & $ 0.203^{+0.003}_{-0.005}$ & $ 78.52^{+1.75}_{-1.79}$ & $ 0.94^{+0.21}_{-0.15}$ & $105^{+6 }_{-11} $      &$ 0.96^{+0.12}_{-0.08} $ & $    81^{+6   }_{-9   }$ & $ 1.99^{+0.10}_{-0.14}$ & $ 0.40^{+0.05}_{-0.07}$ & 285/313  \\
C5& $ 1.75^{+0.04}_{-0.06} $ & $ 10.6^{+0.39}_{-0.44} $ & $ 0.216^{+0.006}_{-0.008}$ & $ 82.45^{+1.92}_{-2.02}$ & $ 0.76^{+0.07}_{-0.09}$ & $102^{+13}_{-18} $      &$ 1.67^{+0.15}_{-0.10} $ & $    47^{+5   }_{-5   }$ & $ 4.77^{+0.54}_{-0.32}$ & $ 0.38^{+0.05}_{-0.06}$ & 305/296  \\
N1& $ 1.63^{+0.03}_{-0.03} $ & $ 5.55^{+0.71}_{-0.58} $ & $ 0.159^{+0.008}_{-0.012}$ & $ 81.37^{+2.48}_{-3.55}$ & $ 2.18^{+0.38}_{-0.28}$ & $321^{+14}_{-13} $      &$ 3.00^{+0.32}_{-0.35} $ & $ 28882^{+1711}_{-1820}$ & $     ^{     }-{     }$ & $     ^{     }-{     }$ & 763/719  \\
N2& $ 1.48^{+0.03}_{-0.05} $ & $ 3.84^{+0.55}_{-0.41} $ & $ 0.132^{+0.003}_{-0.007}$ & $ 81.29^{+1.70}_{-2.23}$ & $ 3.40^{+0.20}_{-0.24}$ & $415^{+15}_{-21} $      &$ 3.55^{+0.33}_{-0.22} $ & $ 28114^{+2987}_{-3145}$ & $     ^{     }-{     }$ & $     ^{     }-{     }$ & 798/711  \\
N3& $ 1.51^{+0.04}_{-0.05} $ & $ 5.06^{+0.14}_{-0.22} $ & $ 0.119^{+0.002}_{-0.008}$ & $ 81.37^{+1.38}_{-1.65}$ & $ 3.03^{+0.19}_{-0.13}$ & $366^{+12}_{-19} $      &$ 3.57^{+0.28}_{-0.21} $ & $ 37812^{+4547}_{-4821}$ & $     ^{     }-{     }$ & $     ^{     }-{     }$ & 812/819  \\

\hline
\end{tabular}
\end{table}
% \end{landscape}

% \begin{landscape}
\begin{table}
\caption{{\tt MYTORUS} model fitted results -- {\tt Decoupled} configuration}
\label{tab:mytd}
\begin{tabular}{lccccccccccccc}
\hline
\hline
ID &$\Gamma $&PL. norm &$N_{H,Z}$ & $N_{H,S}$ &$A_{S}=A_{L}$& EW & $F_{K\alpha}$ & FWHM & $f_s$& $kT^{apec}$ &$L_{2-10 keV, int}$ &$\chi^2$/dof \\
   &   &( $10^{-3}$ ph cm$^{-2}$ s$^{-1}$)& ($10^{24}$ cm$^{-2}$) & ($10^{24}$ cm$^{-2}$) & & (eV)&($10^{-13}$ ergs cm$^2$ s$^{-1}$) &(km s$^{-1}$) &($10^{-2}$) & (keV)& ($10^{41}$ ergs s$^{-1}$) &  \\
\hline
S1 & $1.84^{+0.04}_{-0.03} $ & $ 40.2^{+0.41}_{-0.37} $ & $ 0.218^{+0.004}_{-0.003} $ & $ 0.115^{+0.005}_{-0.005} $ & $ 1.06^{+0.03}_{-0.05} $ & $ 120^{+14}_{-16} $       &$ 1.81^{+0.12}_{-0.14} $ & $  5551^{+682 }_{-711 } $ & $ 1.39^{+0.46}_{-0.41} $ & $ 0.25^{+0.02}_{-0.02} $ & $13.86^{+0.58}_{-0.52} $ &2480/2379 \\ 
C1 & $1.75^{+0.04}_{-0.05} $ & $ 7.65^{+0.67}_{-0.72} $ & $ 0.204^{+0.009}_{-0.006} $ & $ 0.113^{+0.004}_{-0.004} $ & $ 0.97^{+0.11}_{-0.12} $ & $ 109^{+10}_{-12} $       &$ 1.88^{+0.11}_{-0.10} $ & $    72^{+5   }_{-8   } $ & $ 1.25^{+0.15}_{-0.18} $ & $ 0.24^{+0.04}_{-0.03} $ & $ 9.73^{+0.89}_{-0.81} $ &285 /291  \\
C2 & $1.73^{+0.02}_{-0.02} $ & $ 8.84^{+0.45}_{-0.52} $ & $ 0.206^{+0.011}_{-0.009} $ & $ 0.118^{+0.004}_{-0.004} $ & $ 0.79^{+0.14}_{-0.10} $ & $ 100^{+9 }_{-7 } $       &$ 1.27^{+0.10}_{-0.15} $ & $    49^{+4   }_{-3   } $ & $ 4.62^{+0.35}_{-0.38} $ & $ 0.33^{+0.04}_{-0.05} $ & $10.89^{+0.62}_{-0.70} $ &347 /346  \\
C3 & $1.79^{+0.04}_{-0.06} $ & $ 8.07^{+0.32}_{-0.27} $ & $ 0.210^{+0.007}_{-0.008} $ & $ 0.116^{+0.008}_{-0.005} $ & $ 0.75^{+0.10}_{-0.14} $ & $  89^{+14}_{-18} $       &$ 1.87^{+0.17}_{-0.13} $ & $    49^{+6   }_{-6   } $ & $ 1.43^{+0.10}_{-0.15} $ & $ 0.36^{+0.04}_{-0.08} $ & $ 9.88^{+0.74}_{-0.68} $ &285 /290  \\
C4 & $1.79^{+0.04}_{-0.03} $ & $ 8.55^{+0.19}_{-0.23} $ & $ 0.193^{+0.011}_{-0.009} $ & $ 0.117^{+0.004}_{-0.005} $ & $ 0.65^{+0.07}_{-0.09} $ & $ 114^{+8 }_{-10} $       &$ 1.04^{+0.08}_{-0.10} $ & $    82^{+3   }_{-9   } $ & $ 2.04^{+0.13}_{-0.09} $ & $ 0.39^{+0.03}_{-0.04} $ & $11.23^{+0.75}_{-0.81} $ &288 /312  \\
C5 & $1.73^{+0.09}_{-0.07} $ & $ 9.66^{+0.20}_{-0.29} $ & $ 0.211^{+0.011}_{-0.009} $ & $ 0.112^{+0.003}_{-0.003} $ & $ 0.36^{+0.05}_{-0.10} $ & $  95^{+13}_{-7 } $       &$ 1.93^{+0.12}_{-0.08} $ & $    49^{+6   }_{-7   } $ & $ 4.17^{+0.17}_{-0.22} $ & $ 0.45^{+0.07}_{-0.10} $ & $12.84^{+1.22}_{-0.93} $ &355 /311  \\
N1 & $1.62^{+0.03}_{-0.04} $ & $ 6.01^{+0.21}_{-0.32} $ & $ 0.144^{+0.008}_{-0.006} $ & $ 0.112^{+0.003}_{-0.003} $ & $ 2.28^{+0.18}_{-0.21} $ & $ 329^{+15}_{-21} $       &$ 3.05^{+0.23}_{-0.21} $ & $ 30483^{+2114}_{-2214} $ & $     ^{     }-{     } $ & $     ^{     }-{     } $ & $10.66^{+0.28}_{-0.17} $ &747 /716  \\ 
N2 & $1.46^{+0.03}_{-0.02} $ & $ 3.89^{+0.16}_{-0.20} $ & $ 0.123^{+0.007}_{-0.007} $ & $ 0.117^{+0.002}_{-0.003} $ & $ 3.08^{+0.29}_{-0.22} $ & $ 419^{+17}_{-22} $       &$ 3.57^{+0.35}_{-0.41} $ & $ 28407^{+2983}_{-3218} $ & $     ^{     }-{     } $ & $     ^{     }-{     } $ & $ 8.56^{+0.24}_{-0.27} $ &755 /710  \\
N3 & $1.51^{+0.04}_{-0.06} $ & $ 5.63^{+0.19}_{-0.21} $ & $ 0.108^{+0.006}_{-0.005} $ & $ 0.114^{+0.003}_{-0.003} $ & $ 2.61^{+0.17}_{-0.25} $ & $ 368^{+14}_{-19} $       &$ 3.57^{+0.29}_{-0.33} $ & $ 39116^{+3671}_{-4214} $ & $     ^{     }-{     } $ & $     ^{     }-{     } $ & $10.16^{+0.23}_{-0.22} $ &811 /816  \\

\hline
\end{tabular}
\leftline{Errors quoted in Table~\ref{tab:PL}, ~\ref{tab:mytc}, and ~\ref{tab:mytd} are with 90 per cent confidence level.}
\end{table}

\end{landscape}
\newpage

\section{Discussion}
\label{sec:discussion}
We studied NGC~6300 between 2007 and 2016 by using the data from the {\it Suzaku}, {\it Chandra}, and {\it NuSTAR} observations.
We mainly concentrate our study on the nuclear core and circumnuclear torus region. However, there is a possibility that the 
extra-nuclear X-ray sources could contribute to the X-ray emission and affect our analysis. The X-ray luminosity of NGC~6300 is 
$\sim 10^{43}$ ergs s$^{-1}$. Although, the time-averaged luminosity of X-ray binaries can reach reach up to $\sim 10^{37}$ ergs 
s$^{-1}$ \citep{RM2006,Tetarenko2016}. The maximum luminosity of an accretion powered X-ray Pulsar can reach up to $\sim 10^{37}-10^{39}$ 
ergs s$^{-1}$ depending on a range of magnetic field between $\sim 10^{14}-10^{15}$ G \citep{Mushotzky2015}. The ultra-luminous 
X-ray (ULX) sources show luminosity about $< 10^{41}$ ergs s$^{-1}$. The X-ray luminosity of supernovae can reach as high as
$\sim 10^{41}$ ergs $s^{-1}$ in the initial days. Thus, the luminosity of supernovae is $\sim~0.01-1$ per cent of our source of interest 
in this work. Even then, within the period of observations used in present work, no such events were recorded from the optical observations
around the field of view of NGC~6300. Thus, contamination from these sources can be neglected for variability studies.

The positional accuracy of the {\it Suzaku}/XIS is around $19''$ at 90 per cent confidence level \citep{Uchiyama2008}.
At a similar confidence level, {\it Chandra} has $0.8''-2''$ positional accuracy depending on the brightness of the
source \citep{Beckerman2004,Broos2010}. In case of {\it NuSTAR}, the positional accuracy varies between $8''-20''$
\citep{Lansbury2017}. For {\it Suzaku} and {\it NuSTAR}, there can be thousands of transient sources within the source 
region. The number reduces to hundreds for {\it Chandra}. Nevertheless, none of them can be as bright as the AGN core itself. 
Thus, the X-ray emission must originate from the nucleus \citep{Gierlinski2008}.

\label{sec:torus}
\subsection{Properties of torus}
We refer the circumnuclear absorbing material as `torus'. It does not have to be a uniform torus; it could
be a clumpy or patchy torus. We obtained the line-of-sight column density ($N_{H,Z}$) and equatorial column
density ($N_{H,eq}$) from the {\tt powerlaw} and {\tt MYTORUS-coupled} model fitting, respectively. We 
observed that $N_{H,eq}$ is slightly higher than $N_{H,Z}$. This is expected as matter density is expected 
to be high in the equatorial region. Nonetheless, they both show a similar trend over the years. 

During the 2007 epoch, the column density was high with $N_{H,eq} = 2.32 \times 10^{23}$ cm$^{-2}$. It decreased 
over the years and after 9 years, it became almost half with $N_{H,eq} = 1.19 \times 10^{23}$ cm$^{-2}$. The 
overall column density of the absorber is found to be varying over nine years. {\it Chandra} observed the source 
five times within 11 days in 2009. During that epoch, the $N_{H,eq}$ varied between $1.99 \times 10^{23}$ cm$^{-2}$
and $2.16 \times 10^{23}$ cm$^{-2}$. Considering a uniform torus around the AGN, the variation of column density 
within a few days remains a puzzle. However, in an alternative scenario, \citet{Risaliti2002} showed that the 
column density could be changed in timescale as short as a day if the torus is clumpy. To further investigate 
the nature of the torus, we used the {\tt decoupled} configuration of the {\tt MYTORUS} model. We find (see 
Table~\ref{tab:mytd}) that the line-of-sight column density ($N_{H,Z}$) changed over the years while the 
global-averaged column density ($N_{H,S}$) remained almost same. 

The variation of $N_{H,Z}$ can be explained by the migrating clouds in the line of sight. During the {\it Chandra}
observations in 2009, minimum change in the line of sight column density was $\sim 2 \times 10^{21}$ cm$^{-2}$ 
(between C1 \& C2). If we assume that the column density of each cloud is $\sim 2 \times 10^{21}$ cm$^{-2}$, 
then the number of clouds in the line of sight in 2009 would be $102 \pm 10$. Similarly, \citet{Guainazzi2016} 
reported that the number of clouds in the line of sight for Mkn~3 was $17 \pm 5$ in 2014. NGC~6300 is known
as `changing look' AGN \citep{Matt2003,Guainazzi2002}. \cite{Guainazzi2002} argued that the temporary `shut-off' 
of the nucleus could be the reason for the `changing-look'. However, the transiting clouds in the line of sight 
would naturally explain the variation of observed flux and the line of sight column density \citep{Yaqoob2015}. 
The observed flux depends on both lines of sight column density and nuclear activity. Thus, the observed flux
can be changed due to the transiting clouds in the line-of-sight even if the nucleus remains unchanged. 
Therefore, the `changing look' AGN can be explained naturally by the transition of clouds in the line of sight. 

Another possible explanation for $N_{H}$ variation could be the ionization of the `torus'. The outer surface of
the torus can be ionized by the radiation from the AGN that has leaked through the clumpy torus \citep{Guainazzi2016}.
\citet{LaMassa2015} showed that the transition of changing look quasar SDSS~J015957.64+003310.5 could be explained by the 
ionization, but not with the transiting clouds. Nevertheless, in the case of NGC~6300, no ionized line was observed. Thus, it 
is improbable that the variable column density is due to ionization.

\label{sec:nuclear}
\subsection{Nuclear Emission}

We find the variability above 3 keV for NGC~6300 in all nine observations. The calculated rms fractional 
variability amplitude in $3-10$~keV or higher ($10-40$~keV) energy bands for all the observations are 
presented in Table~\ref{tab:fvar}. The observed variability of the primary emission (>3 keV) is consistent 
with the other Seyfert~2 galaxies \citep{HG2015}.

In the case of supermassive black holes, the X-ray emission is mostly originated from the Compton cloud. 
The properties of the X-ray spectra (popularly charactarized with the photon index $\Gamma$) depend 
on the properties of the Compton cloud which is charactarized by hot electron temperature ($kT_e$), 
optical depth ($\tau$) and cut-off energy ($E_c$). We obtained $\Gamma$ from the {\tt powerlaw} model, 
$kT_e$ and $\tau$ from the {\tt compTT} model (see Table~\ref{tab:PL}). We tried to estimate $E_{cut}$ 
using the {\tt cutoffpl} model, but were unable to constrain it. All observations indicated
$E_c > 500$~keV which is the upper limit of the {\tt cutoffpl} model. The {\tt highecut} model also 
failed to constrain $E_{cut}$ of the source. Nevertheless, considering an isotropic electron cloud with 
seed photon source placed at the centre, as sugested by \cite{P2001}, one can get an estimate of cutoff
energy using $E_c\sim 3kT_e$ for $\tau \gg 1$, and $E_c \sim 2kT_e$ for $\tau \leq 1$. Using the relation
and the parameters obtained from the {\tt compTT} model, it can be implied that the cut-off energy might be 
$60.0 < E_c < 200.0$ keV and rise up neatly within this period of time. We also noticed the increase 
in $E_c$ with the time considering the simple theoretical approach. However, it would be naive to draw
any conclusion based on the cut-off energy variation. The cut-off energy ($E_c$) could not be constrained 
\citep{Leighly99,Guainazzi2002} by earlier observations too.

\begin{figure*}
\vskip 0.5cm
\includegraphics[width=16cm]{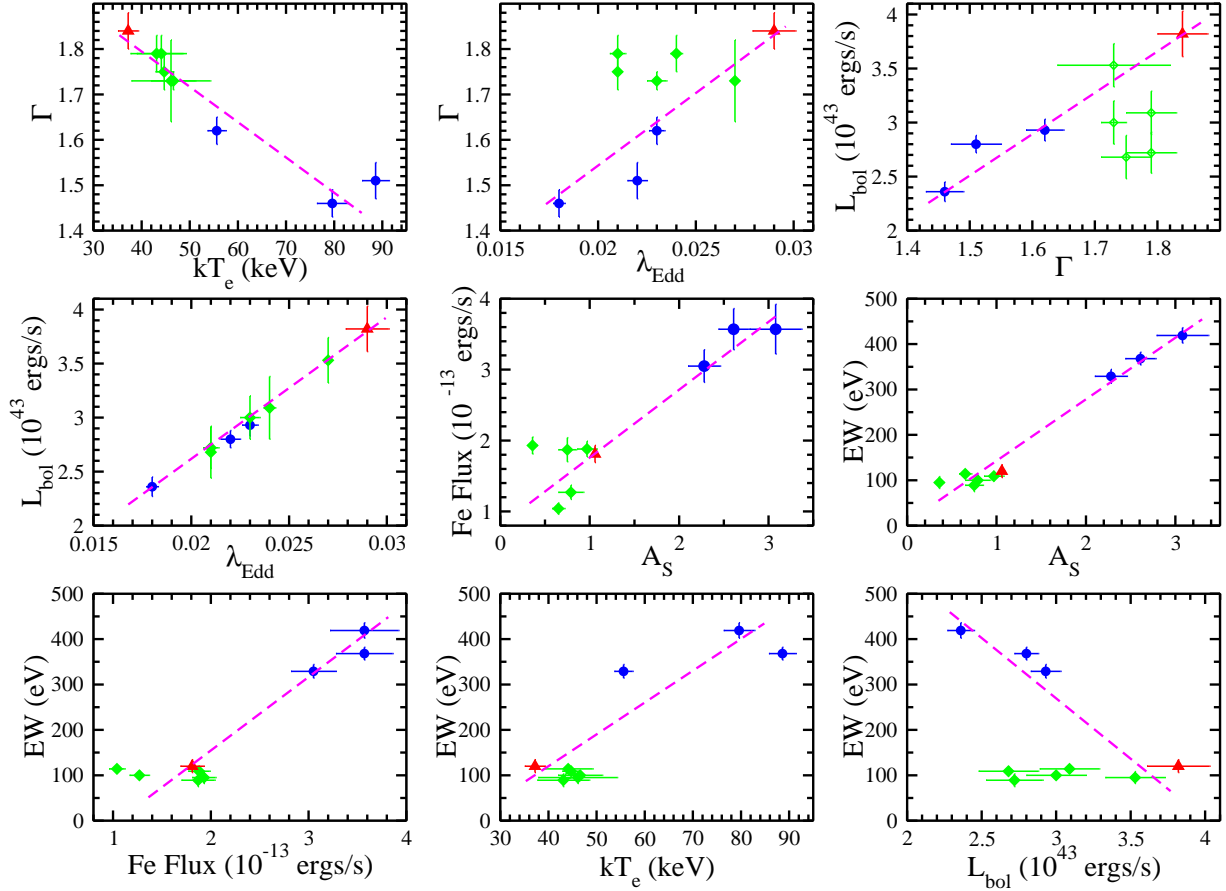}
\caption{Correlation plots of numerous spectral fitted parameters of NGC~6300. The red, green, and blue points
are for the {\it Suzaku, Chandra} and {\it NuSTAR} observations, respectively. Note the short term variation
of $L_{bol}$ and $\lambda_{Edd}$ during the {\it Chandra} observations.}
\label{fig:6}
\end{figure*}

In our observation, we found $\Gamma \sim 1.7-1.8$ during the 2007 \& 2009 epochs. During these epochs, 
the Compton cloud temperature was $\sim 40$~keV. Later, in 2013 epoch, $kT_e$ was found to increase, whereas
the value of $\Gamma$ decreased. This is expected since the hotter Compton cloud emits harder and flatter spectrum. 
We found an anti-correlation between $kT_e$ and $\Gamma$ with Pearson correlation index of -0.9187. This strong 
anti-correlation indicates that the hot electron cloud is the primary source for the harder spectra. The 
correlation is presented in Figure~\ref{fig:6}.

We calculated source intrinsic luminosity ($L_{in}$) in $2-10$~keV energy range and found to be varied over 
the years. $L_{in}$ was $1.4 \times 10^{42}$ ergs s$^{-1}$ in 2007. After that, it decreased in later epochs. $L_{in}$ 
varied between $0.86 \times 10^{42}$ ergs s$^{-1}$ and $1.18 \times 10^{42}$ ergs s$^{-1}$ in 2009, 2013 \& 2016 epochs. The 
variation of accretion rate could be responsible for the variation of $L_{in}$. To get an estimate of the accretion 
rate, we calculated the bolometric luminosity with bolometric correction $\kappa_{bol} = 1.44 \pm 0.12$ dex 
\citep{Brightman2017}. We estimated the Eddington luminosity of the source as $1.3 \times 10^{45}$ ergs s$^{-1}$ 
assuming the mass of the BH as $10^7$ $M_{\odot}$. The bolometric luminosity depends on the mass accretion 
rate as, $L_{bol} = \eta \dot{M} c^2$, where $\eta$ is the energy conversion efficiency. The Eddington 
luminosity is $L_{Edd} = \eta \dot{M}_{Edd} c^2$. Thus, the mass accretion rate ($\dot{M}$) in terms of 
Eddington mass accretion rate ($\dot{M}_{Edd}$) is, $\dot{m} = \dot{M}/\dot{M}_{Edd}$. The Eddington ratio 
($\lambda_{Edd}$) is defined as $\lambda_{Edd}=L_{bol}/L_{Edd}$. Since $\eta$ is the energy conversion efficiency, 
$L_{Edd} = \eta \dot{M}_{Edd} c^2$ and $L_{bol} = \eta \dot{M} c^2$. Therfore, $L_{bol}/L_{Edd} = \lambda_{Edd} = 
\dot{m}$. The observed $\dot{m}$ or the Eddington ratio is consistent with other Seyfert~2 galaxies 
\citep{WL2004, SSL2007}.

We found that the intrinsic luminosity ($L_{in}$) is strongly correlated with the Eddington ratio ($\lambda_{Edd}$)
with the Pearson correlation index as 0.9967. We also found a strong correlation between $\lambda_{Edd}$ and $\Gamma$ 
with Pearson correction index of 0.7472. The correlation of $\lambda_{Edd}$ and $\Gamma$ is studied by many authors
\citep{Lu1999,Shemmer2006,Shemmer2008,Risaliti2009} and a positive correlation is found for NGC~6300 as well. 
We also found a correlation between $L_{in}$ and $\Gamma$ with Pearson correlation index as 0.7604. This positive
correlation of $\Gamma$ and $L_{in}$ is observed in high redshift objects \citep{Dai2004,Saez2008}, but not in 
the local universe \citep{Brightman2011}.

Above three correlations ($L_{in}-\lambda_{Edd}$, $\lambda_{Edd}-\Gamma$, $L_{in}-\Gamma$) can be explained 
in a single framework of variable accretion rate. If the mass accretion rate increases, the radiated energy 
will increase, thus, the intrinsic luminosity. Hence, $\dot{m} \propto L_{in}$, or $\lambda_{Edd} \propto L_{in}$. 
Again, if the luminosity increases, it will cool down the Compton cloud more efficiently implying a steeper photon
index, i.e. $L_{in} \propto \Gamma$. These relations are well known for Galactic black holes \citep{RM2006}.

\label{sec:reflection}

\subsection{Reflection \& Line Emission}
Hard X-ray emission from the Compton cloud is reflected from a cold neutral medium surrounding the AGN \citep{GF1991,Matt1991}. 
Within our spectral domain, the reflection component consists of a reflection hump and an iron fluorescent emission line. 
Due to higher inclination, the Seyfert~2 galaxies usually show stronger reflection than the Seyfert~1 galaxies \citep{Ricci2011}. 
The strength of the reflection component can be obtained from the {\tt pexrav} ($R$, relative reflection) or {\tt MYTORUS}
($A_S$; relative normalization of scattering) model. As the reflection hump is observed in $15-30$~keV, $R$ may not be
constrained with the {\it Chandra} data. In this case, the {\tt MYTORUS} model can constrain $A_S$ as both the reflected 
and the line emission are computed self-consistently.

Along with the reflection hump, the iron line emission is a good indicator of reflection. NGC~6300 is known to emit 
strong reflection component with unusually high $R$ (> 4) \citep{Leighly99}. The reflection dominated {\it RXTE} 
spectrum was expected to show EW $\sim 1$~keV although the observed EW was about $\sim 470$~eV. \citet{Leighly99} argued 
that the sub-solar abundances were the reason behind the observation of narrow line-width than expected. The EW of Fe 
fluorescent line was decreased in 1999 and 2001 observations with 140 eV and 69 eV, respectively 
\citep{Guainazzi2002,Matsumoto2004} although the relative reflection was still high.

In our analysis, the Fe K$\alpha$ line was observed in the spectra of NGC~6300 during 2007, 2013 \& 2016 epochs from the 
{\tt powerlaw} model fitting. However, we did not detect the Fe K$\alpha$ line in the spectra from the 2009 epochs 
while fitting the data with the {\tt powerlaw} model. On the other hand, the {\tt MYTORUS} model fitting result showed the 
presence of Fe line in all observations. This discrepancy arose as the {\tt MYTORUS} model compute Fe line self-consistently,
while an ad-hoc {\tt Gaussian} component was added as Fe line in the {\tt powerlaw} model. Thus, we made our discussion on
the reflection and Fe K$\alpha$ line based on the results obtained from the spectral analysis with {\tt MYTORUS} model. 
During 2007, the iron K$\alpha$ line was observed with an equivalent width of $120^{+14}_{-16}$ eV (see Table~\ref{tab:mytd}). 
In 2009 epoch, a marginally narrower line (95-114 eV) was observed from the {\it Chandra} observations. Though similar EW 
was observed in 2007 \& 2009 epochs, a different relative scattering normalization was found. In 2007, $A_S = 1.06$ indicates
that the reflection dominates over the primary continuum. In 2009, the reflection was weaker compared to 2007 ($A_S <1$).
However, in 2013 \& 2016 epochs, the reflection became stronger ($A_S > 2$ with broad EW (>300 eV). The observed EW is 
consistent for the reflection from a Compton thin reprocessor \citep{Matt2003} for all the observations.

We found that the EW and $A_S$ are strongly correlated with Pearson correlation index of 0.9893. This indicates that 
the EW strongly depends on the strength of the reflection. The Fe K$\alpha$ line flux is found to be strongly correlated 
with the EW and $A_S$ with Pearson correlation indices of 0.9347 and 0.9275, respectively. This suggests that the Fe-line 
flux strongly depends on the strength of the reflection. The EW also correlates with the Compton cloud temperature 
($kT_e$) and has a correlation value of 0.9146. From this, one can implicate that the hotter Compton cloud causes more 
reflection from various regions which broadens the line-width. However, the Pearson correlation index between 
Fe K$\alpha$ line flux and $L_{bol}$ is estimated to be -0.5187, which suggests a moderate anti-correlation. This 
X-ray Baldwin effect was observed in several other sources \citep{Iwasawa1993, Ricci2013a, Boorman2018}. Nonetheless, 
the 2009 epoch seems to exhibit outliers of the Baldwin effect for NGC~6300. The reason behind this could be a separate 
origin of the iron line than the other epochs. 

Later, we focus on the Fe K$\alpha$ line emitting region. We calculated the Fe K$\alpha$ line emitting region from
observed FWHM. In 2007, the FWHM of the Fe K-line was derived to be 5550 km s$^{-1}$ which indicates that the line 
emitting region would be $\sim 3900$ $r_g$ away from the central source, i.e. the broad-line region (BLR). In 2009, 
we observed a narrow FWHM (< 100 km s$^{-1}$), which means the line emitting region is at $>10^7$ $r_g$, i.e. $>17$ pc 
located near the `torus' region. In 2013, the {\it NuSTAR} observations revealed a broad FWHM with $\sim$ 30000 km 
s$^{-1}$. It implies that the line emitting region is $\sim 130$ $r_g$ away. In Jan 2016 \& Aug 2016 observations, 
the values of FWHM were $\sim 28000$ km s$^{-1}$ and $39000$ km s$^{-1}$, respectively. In these two epochs also, 
the line emitting region would be $\sim 150$ $r_g$ and $\sim 80$ $r_g$ away, respectively. It is believed that the 
narrow Fe line is ubiquitous and can be emitted from the `torus', the BLR or the accretion disc \citep{Nandra2006}. 
From our analysis of NGC~6300, we found that the Fe fluorescent line was emitted from separate regions during various 
epochs. \citet{Guainazzi2002} estimated the Fe K$\alpha$ line emitting region for NGC~6300 to be $\sim 10^4$ $r_g$ in 
2001, i.e. torus. During 2007, we find that the Fe-line was emitted from the BLR region. In 2009 epoch, we observed that
the line emitting region as the `torus'. But, during 2013 and 2016 epochs, the iron line was emitted from the accretion
disc. It is also possible that the narrow Fe K$\alpha$ line was emitted from the `torus' in 2013 and 2016 epochs, but,
could not be detected due to the presence of broader Fe K$\alpha$ line emission from the accretion disc. 

Considering the time-delay patterns between Fe-line and continuum during 2013 and 2016 epochs (see Figure~\ref{zdcf}), 
one can get a rough estimate of the size of the Compton cloud. Since the delay between these two is minimal 
compared to an AGN, it is possible that the Comptonized and reflected Fe-line component originated from a similar 
vicinity. Bearing this in mind, the broad iron line emitting region could be the farthest extents of Compton cloud
during 2013 and 2016 epochs.    

\begin{table*}
\centering
\caption{Evolution of line-of-sight column density, luminosity and accretion rate is
shown from Feb 1997 to August 2016.}
\label{tab:lum}
\begin{tabular}{lcccccc}
\hline
ID &Date&$N_{H,Z}$ &$\Gamma$&$L_{bol}$ & $\lambda_{Edd}$ \\
 &(MJD)&($10^{23}$ cm$^{-2}$)& &($10^{43}$ ergs s$^{-1}$) & \\ 
\hline
R$^1$& 50493.50 &$5.80\pm0.22 $&$1.71\pm0.20$   & $0.69^*$& $0.005^* $   \\  
B$^2$& 51418.50 &$2.10\pm0.10 $&$2.19\pm0.10$   & $5.23^*$& $0.041^* $   \\ 
X$^3$& 51971.50 &$2.40\pm0.15 $&$1.94\pm0.09$   &   - &  -            \\ 
S1   & 54390.51 &$2.18\pm0.04 $&$1.84\pm0.04$   & $3.82\pm0.21$& $0.029\pm0.001 $   \\ 
C1   & 54985.27 &$2.04\pm0.08 $&$1.75\pm0.03$   & $2.68\pm0.20$& $0.021\pm0.001 $   \\ 
C2   & 54989.17 &$2.06\pm0.10 $&$1.73\pm0.02$   & $3.00\pm0.20$& $0.023\pm0.001 $   \\ 
C3   & 54991.01 &$2.10\pm0.08 $&$1.79\pm0.05$   & $2.72\pm0.19$& $0.021\pm0.001 $   \\  
C4   & 54992.88 &$1.93\pm0.10 $&$1.79\pm0.04$   & $3.09\pm0.20$& $0.024\pm0.001 $   \\ 
C5   & 54996.33 &$2.11\pm0.10 $&$1.73\pm0.08$   & $3.53\pm0.20$& $0.027\pm0.001 $   \\ 
N1   & 56348.89 &$1.44\pm0.07 $&$1.62\pm0.04$   & $2.93\pm0.10$& $0.023\pm0.001 $   \\ 
N2   & 57411.03 &$1.23\pm0.07 $&$1.46\pm0.02$   & $2.36\pm0.09$& $0.018\pm0.001 $   \\ 
N3   & 57624.35 &$1.08\pm0.06 $&$1.51\pm0.05$   & $2.80\pm0.08$& $0.022\pm0.001 $   \\ 
\hline
\end{tabular}
\leftline{$^1$ RXTE observation \citep{Leighly99}, $^2$ BeppoSAX observation \citep{Guainazzi2002},}
\leftline{$^3$ XMM-Newton observation \citep{Matsumoto2004}.} \\
\leftline{$^*$ Error is not quoted.}
\end{table*}

\label{sec:soft}
\subsection{Soft Excess}

Soft excess ($<$ 3 keV) is found almost in every AGN. However, the origin of the soft excess is poorly understood. 
One possible origin of soft excess could be the reflection from an optically thick warm Comptonizing region 
\citep{Gierlinski2004,Magdziarz1998} or the reflection from the ionized accretion disc \citep{Fabian2002,
Ross2005,Walton2013}. The origin of soft excess can be explained by the heating of circumnuclear gas from the shock 
produced by AGN outflows \citep{King2005} or photoexcitation and photoionization of circumnuclear gas of the primary 
emission of the AGN. The high-resolution capabilities of X-ray observatories such as {\it XMM-Newton} and {\it Chandra} 
have supported the later scenario in recent studies, e.g. NGC~1068 \citep{Young2001}; \citep{Kinkhabwala2002}, the 
Circinus galaxy \citep{Sambruna2001}, Mkn~3 \citep{Sako2000}; \citep{Bianchi2005b}; \citep{Pounds2005}, \citep{Guainazi2007}. 
\citet{Fukumura2016} proposed shock heating near the ISCO could produce the soft excess. The theory successfully
demonstrated the spectra of Seyfert 1 galaxy Ark~120. \citet{Kaufman2017} presented Bulk Motion Comptonization as a 
plausible cause of the soft excess. \citet{Done2012} presented a new perspective on the soft excess by attaching the 
component with the high mass accretion rate of the disc itself.

In our observations of NGC~6300, we modelled the soft excess with {\tt constant*(powerlaw + apec)}. The {\tt powerlaw}
normalization and $\Gamma$ were tied with the corresponding parameters of the primary component. We find that the 
soft excess contributed about up to $\sim 6$ per cent. The {\tt APEC} model temperature varied between 0.2 and 0.5 keV, 
indicating a warm absorber. We also checked the fractional variability rms amplitude ($F_{var}$) for the soft excess, 
i.e. in the energy band of $0.5-3$~keV. We found variability in this energy band in some observations (see 
Table~\ref{tab:fvar}). The average variability is higher in the soft excess than the primary emission. 
\citet{Done2012} claimed that for lower $L/L_{Edd}$ sources, the energy-dependent variabilities 
is less for the soft excess part. However, this contradicts our findings. Higher variabilities in the soft 
excess indicate that the soft excess could be the scattered primary emission from the warm reflector or 
accretion disc. The absence of variability in some observations infers that it might have originated from 
the ambient medium during those observations. From the spectral studies (see Table~\ref{tab:PL}, \ref{tab:mytd}), 
we also found variable $f_s$ which implicates a dynamic mechanism such as reflection or complex absorbing medium 
\citep{SD2007} could be responsible for the origin of the soft excess. Overall, the origin of the soft excess 
is complex. It is plausible that more than one factor contributed to the soft excess part of the spectra.

\begin{figure}
\includegraphics[width=\columnwidth]{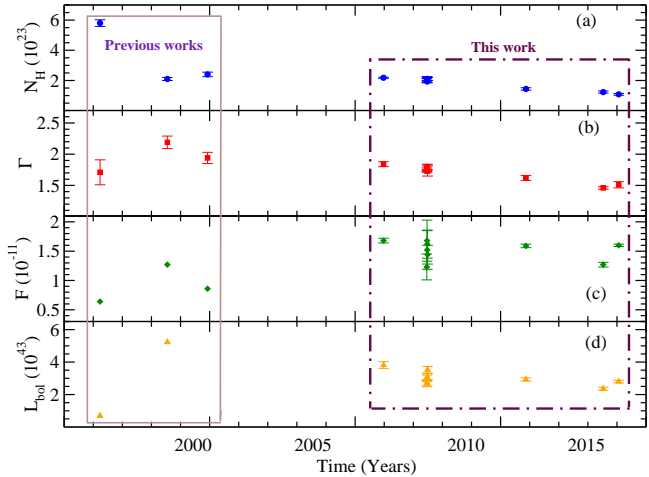}
\caption{Variation of (a) line-of-sight column density ($N_H$) in 10$^{23}$ cm$^{-2}$ unit, 
(b) photon index ($\Gamma$), (c) $2-10$~keV absorbed flux ($F$) in 10$^{-11}$ ergs cm$^{-2}$ s$^{-1}$ unit
and (d) $2-10$~keV bolometric luminosity ($L_{bol}$) in 10$^{43}$ ergs s$^{-1}$ unit are shown over the years.}
\label{fig:past}
\end{figure}

\label{sec:evolution}
\subsection{Evolution of the System}

We studied the Seyfert~2 galaxy NGC~6300 between 2007 and 2016. During the period of observation, the source
evolved over the years. The source was observed in the Compton-thick region in 1997 \citep{Leighly99}, 
though, it was observed in the Compton-thin region in 1999 \citep{Guainazzi2002}. This made NGC~6300 a changing-look
AGN \citep{Matt2003}. However, during our observation period between 2007 and 2016, NGC~6300 remained Compton-thin 
(as $N_H < 1.5 \times 10^{24}$ cm$^{-2}$). Although, we observed a change in the line-of-sight column density ($N_{H,Z}$)
over the years, the global averaged column density ($N_{H,S}$) remained constant. It can be explained with 
the transiting clouds along the line-of-sight (see section~\ref{sec:torus}). In Figure~\ref{fig:past} (a), 
we show the evolution of the line-of-sight column density over the years.

In addition to the evolution of the circumnuclear properties, the nuclear region also evolved over the years
(see Table~\ref{tab:lum}). The photon index ($\Gamma$), absorbed flux, and luminosity changed over the years. 
In 1997, the source was observed with very low luminosity with bolometric luminosity, $L_{bol} =6.9 \times 10^{42}$ 
ergs s$^{-1}$. In 1999, the bolometric luminosity increased to $L_{bol} = 5.23 \times 10^{43}$ ergs s$^{-1}$. The photon 
index also increased to $\Gamma = 2.11$ from $\Gamma = 1.71$ in 1999. Since then, the photon index decreased over the 
years till 2016. However, it would be naive to conclude that the $\Gamma$ decreased gradually since the source 
was not observed on a regular basis. Nonetheless, both the flux and luminosity changed over the years. The change in 
the luminosity can be explained with the evolution of the mass accretion rate. We also checked for a correlation between
the intrinsic luminosity $L_{in}$ and the line-of-sight column density ($N_{H,Z}$) which we fail to observe. Such a 
correlation would indicate luminosity dependent covering factor \citep{Ricca2013b}. Since no correlation was found 
between these two, it is likely that the `torus' and the AGN evolved independently.

\section{Conclusion}
\label{sec:conclusion}

We study NGC~6300 between 2007 \& 2016. Over the 9 years of observations, we investigated
and found that NGC~6300 evolves with time. NGC~6300 was previously reported as changing-look 
AGN. However, we find it in Compton-thin region in every observation. Following are the findings 
from our work.

\begin{enumerate}
\item[1.] The obscured torus is not uniform; rather, it is clumpy. Global averaged column
density is constant over the years. However, the line-of-sight column density changes with time. 
This change is interpreted as due to the transiting clouds along the line-of-sight.

\item[2.] The nuclear region was found to evolve over the years. The intrinsic luminosity of the 
source changes with time. The change in the mass accretion rate is likely to be responsible for that.

\item[3.] The torus and primary nucleus evolved independently, at least during our observation. 
We did not find any relation between column density and intrinsic luminosity.

\item[4.] The Fe K$\alpha$ line emitting region is different in different epochs. During 2007, 2009, 
and 2013-16, the primary source of Fe K$\alpha$ emission was BLR, `torus' and accretion disc, 
respectively. Narrow Fe K$\alpha$ line is originated in the torus and could be present in 
all epochs. Although, narrow Fe K$\alpha$ lines were not detected in every epoch in presence of broad Fe K$\alpha$ line.

\item[5.] We find variability in both soft excess (0.5-3 keV range) and primary emission (> 3 keV).
The variability in the soft excess infers that it could be scattered primary emission. However,
a lack of variability in some observations infers that the origin of the soft excess is complex.

\end{enumerate}

\section*{Acknowledgements}
We acknowledge the anonymous Reviewer for the constructive review which improved the clarity of the manuscript.
A.J. and N. K. acknowledges support from the research fellowship from Physical Research Laboratory, Ahmedabad, India,
funded by the Department of Space, Government of India for this work. AC acknowledges Post-doctoral fellowship of S. N. 
Bose National Centre for Basic Sciences, Kolkata India, funded by Department of Science and Technology (DST), India. 
PN acknowledges Council of Scientific and Industrial Research (CSIR) fellowship for this work. This research has made 
use of data and/or software provided by the High Energy Astrophysics Science Archive Research Center (HEASARC), which 
is a service of the Astrophysics Science Division at NASA/GSFC and the High Energy Astrophysics Division of the Smithsonian 
Astrophysical Observatory. This work has made use of data obtained from the {\it Suzaku}, a collaborative mission between 
the space agencies of Japan (JAXA) and the USA (NASA). The scientific results reported in this article are in part based 
on observations made by the Chandra X-ray Observatory. This research has made use of software provided by the Chandra 
X-ray Center (CXC) in the application packages CIAO, ChIPS, and Sherpa. This work has made use of data obtained from the 
{\it NuSTAR} mission, a projects led by Caltech, funded by NASA and managed by NASA/JPL, and has utilised the NuSTARDAS 
software package, jointly developed by the ASDC, Italy and Caltech, USA. This research has made use of the NASA/IPAC 
Extragalactic Database (NED) which is operated by the Jet Propulsion Laboratory, California Institute of Technology, 
under contract with the National Aeronautics and Space Administration. This research has made use of the SIMBAD database, 
operated at CDS, Strasbourg, France.

\section*{Data Availability}
We have used archival data for our analysis in this manuscript. All the models and software used in this manuscript are 
publicly available. Appropriate links are given in the manuscript.

%%%%%%%%%%%%%%%%%%%%%%%%%%%%%%%%%%%%%%%%%%%%%%%%%%



%\bibliographystyle{mnras}


\newpage

%%%%%%%%%%%%%%%%%%%%%%%%%%%%%%%%%%%%%%%%%%%%%%%%%%

\label{lastpage}
\end{document}